\documentclass[10pt,journal,compsoc]{IEEEtran}

\usepackage[pdftex]{graphicx}
\DeclareGraphicsExtensions{.pdf,.jpeg,.png}

\usepackage{microtype}                 % use micro-typography (slightly more compact, better to read)
\PassOptionsToPackage{warn}{textcomp}  % to address font issues with \textrightarrow
\usepackage{textcomp}
\usepackage{mathptmx}                  % use matching math font
\usepackage{times}                     % we use Times as the main font
\usepackage{cite}                      % needed to automatically sort the references
\usepackage{tabu}                      % only used for the table example
\usepackage{booktabs}                  % only used for the table example
\usepackage{subcaption}
\usepackage{comment}

\usepackage{tikz}
\tikzset{>=latex}

\usepackage{amsmath,amsfonts,amssymb}

\usepackage[normalem]{ulem}

\newcommand{\mb}{\mathbf}

\begin{document}

%% Paper title.
\title{Visualizing Time-Varying Particle Flows with Diffusion Geometry}

%% This is how authors are specified in the journal style

%% indicate IEEE Member or Student Member in form indicated below
\author{Matthew~Berger
        and~Joshua~A.~Levine% <-this % stops a space
\IEEEcompsocitemizethanks{\IEEEcompsocthanksitem M. Berger and J. A. Levine are with the Department
of Computer Science, University of Arizona\protect\\
E-mail: \{matthewberger , josh\}@email.arizona.edu}}%

\markboth{Journal of \LaTeX\ Class Files,~Vol.~14, No.~8, August~2015}%
{Berger \& Levine: Visualizing Time-Varying Particle Flows with Diffusion Geometry}

\IEEEtitleabstractindextext{%
\begin{abstract}
	% task and challenge
	The tasks of identifying separation structures and clusters in flow data are fundamental to flow visualization.
	Significant work has been devoted to these tasks in flow represented by vector fields, but there are unique challenges in addressing these tasks for time-varying particle data.
	The unstructured nature of particle data, nonuniform and sparse sampling, and the inability to access arbitrary particles in space-time make it difficult to define
	separation and clustering for particle data.
	% big idea observation
	We observe that weaker notions of separation and clustering through continuous measures of these structures are meaningful when coupled with user exploration.
	We achieve this goal by defining a measure of \emph{particle similarity} between pairs of particles.
	More specifically, separation occurs when spatially-localized particles are dissimilar, while clustering is characterized by sets of particles that are similar to one another.
	% detailed contributions
	To be robust to imperfections in sampling we use \emph{diffusion geometry} to compute particle similarity.
	Diffusion geometry is parameterized by a \emph{scale} that allows a user to explore separation and clustering in a continuous manner.
	We illustrate the benefits of our technique on a variety of 2D and 3D flow datasets, from particles integrated in fluid simulations based on time-varying vector fields,
	to particle-based simulations in astrophysics.
\end{abstract}

\begin{IEEEkeywords}
Computer Society, IEEE, IEEEtran, journal, \LaTeX, paper, template.
\end{IEEEkeywords}}

\maketitle

\begin{figure*}[!t]
	\begin{center}
	\begin{subfigure}[!t]{0.48\textwidth}
		\includegraphics[width=\linewidth]{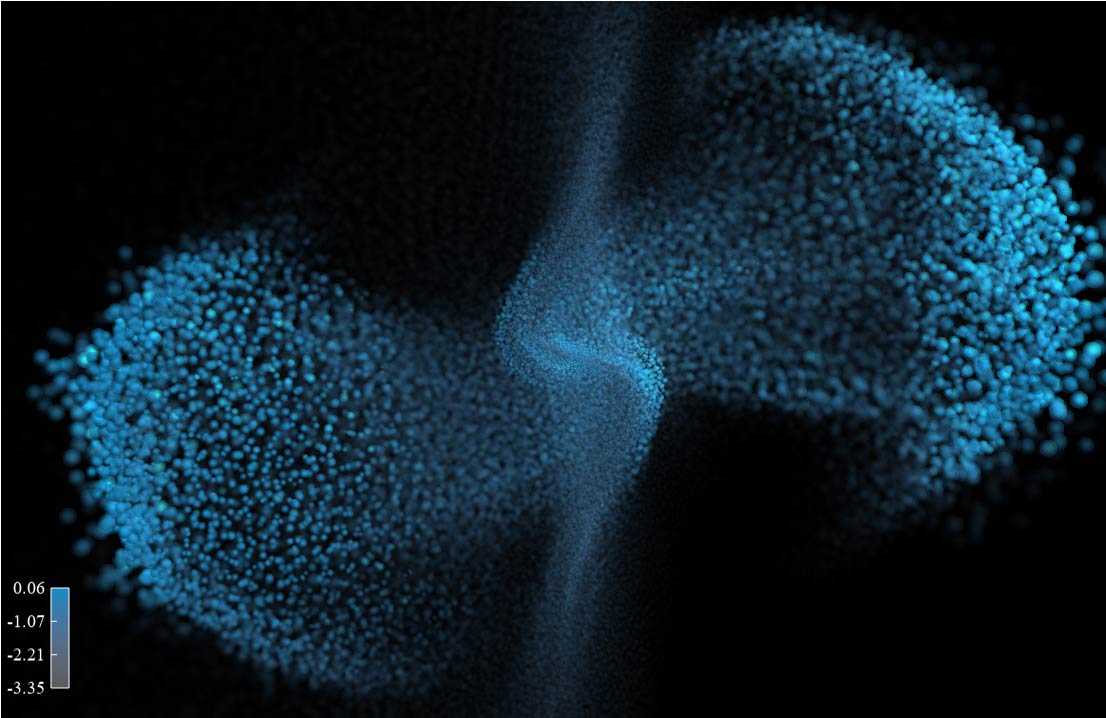}
		\caption{Particle Attraction}
	\end{subfigure}
	\begin{subfigure}[!t]{0.48\textwidth}
		\includegraphics[width=\linewidth]{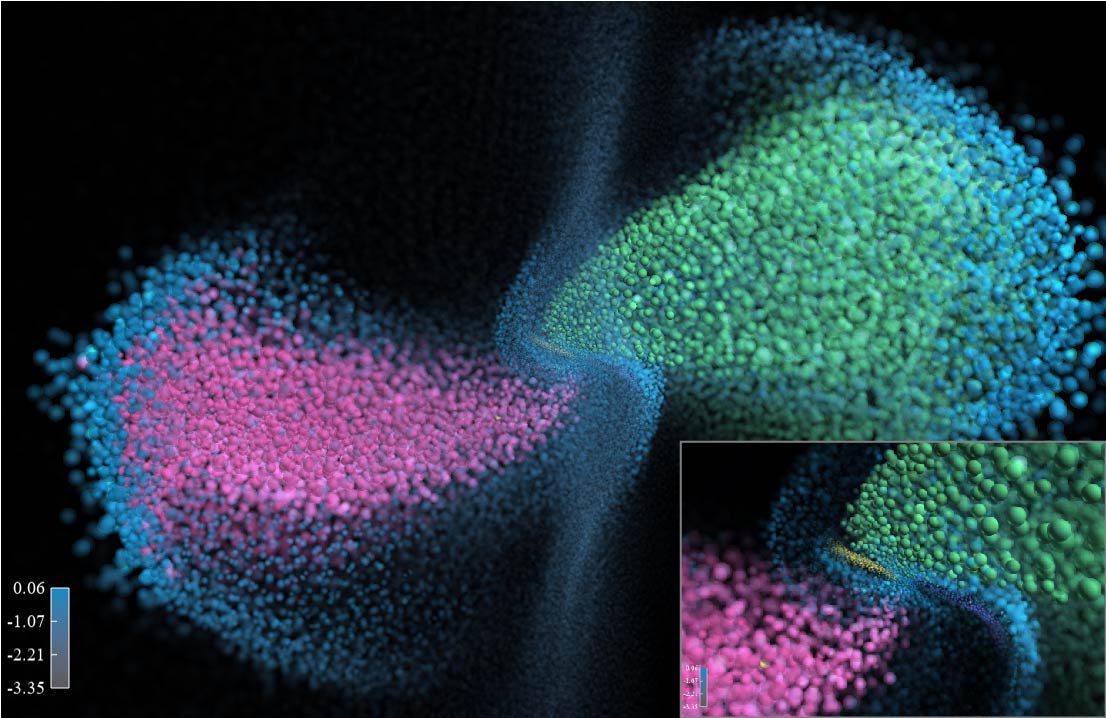}
		\caption{Particle Clustering}
	\end{subfigure}
	\end{center}
  \caption{Our method for computing particle similarities enables the study of attraction (a), equivalent to backward-time separation, and clustering (b) in particle flows, shown here for
  the collapse of a molecular cloud core to a protostar~\cite{lewis2015smoothed} in which the magnetic field is initially misaligned with the rotation axis
  by 45 degrees. Particle attraction (a) reveals a warped disc structure in the center, known to prevent collimated
  outflow~\cite{lewis2015smoothed} and inhibit star formation. Particle similarities (b) reveal broad bipolar outflows as 2 clusters of green and pink particles,
  while the inset shows additional clusters in yellow and purple, further evidence in the disruption of collimated outflow.}
	\label{fig:teaser}
\end{figure*}

\IEEEdisplaynontitleabstractindextext

\IEEEpeerreviewmaketitle

\IEEEraisesectionheading{\section{Introduction}}

% generalities in unsteady flow vis
\IEEEPARstart{A}{nalyzing} and visualizing flow data are vital tasks across a wide range of engineering and scientific disciplines, such as computational fluid dynamics~\cite{OttoKET12},
climate modeling~\cite{janicke2008automatic}, and blood flow~\cite{oeltze2016cluster}. Essential to the understanding
of these physical phenomena is the analysis of flow structures that change over time. For instance, tracking of vortex cores helps analyze
anatomical structures in blood flow~\cite{van2012visualization}, while identifying regions of flow separation relates to drag
on vehicles~\cite{garth2007efficient}. The exploration of these types of structures helps analysts model, understand, and compare flow.

% data representation
The data representation of flow plays a vital role in this type of analysis. Time-varying flow is often
represented by a collection of vector fields that sample a sequence of time steps. Traditional approaches for flow visualization of time-varying vector fields
make a common assumption that the analysis has access to any vector value at any point in space-time. For instance, in the visualization of separating structures,
characterized by repelling or attracting behavior, a common approach is to differentiate the \emph{flow map}, which encodes the destinations of flow integrated across time spans. This is
traditionally done in the computation of the Finite-Time Lyapunov Exponent (FTLE)~\cite{haller2000lagrangian,shadden2005definition}.
Numerical differentiation requires the integration of the vector field at arbitrary points in space and time.
Another common task is flow clustering~\cite{hong2014flda,pobitzer2012statistics}, where the ability to integrate the vector field ensures that one has a representative flow sampling.

% particle data is HARD
Unstructured particle data is being used with increasing prominence in flow simulation as an alternative to vector fields. This resurgence can be seen in cosmology, and in particular
magnetohydrodynamics, where Smoothed Particle Hydrodynamics (SPH) simulations are frequently performed to model the evolution of gas dynamics~\cite{price2012smoothed}.
Additionally, mixed representations of particles and adaptive mesh refinement (AMR) are used for
simulating dark matter in the universe~\cite{wu2012hierarchical}, ultimately resulting in particle data. In analyzing and visualizing particle data, we no longer have
the ability to arbitrarily sample flow trajectories as we do in vector fields.  While there are some approaches to interpolating flow trajectories~\cite{ChandlerOJ15} from particles, these can create errors near highly divergent regions~\cite{CBJ16}. Moreover, particle data can be sparse and nonuniformly sampled in space and time, posing challenges for scattered data interpolation. These characteristics make traditional tasks
in vector field visualization, such as separation and clustering of flows, far more challenging to address.

% observation: similarity
To handle particle data we introduce weaker notions of separation and clustering based on continuous measures of these structures.
Coupled with user exploration, such continuous measures enable the user to assess the strength of separation and clustering present in flows.
The continuous measures of these concepts are based on our observation that the notion of \emph{particle similarity} underlies these traditional flow visualization tasks. More specifically, for measuring flow separation we can analyze the dissimilarity of particles in a local spatial neighborhood.
In clustering flow, similarity plays a key role as it is often a prerequisite for a clustering algorithm. Hence, once we define an appropriate similarity
measure for particles, we can then perform these standard flow visualization tasks for particle data.

% multi-scale similarity: diffusion geometry
Nevertheless, particle similarity still faces the problems inherent in sampling.
To handle these challenges we use the methodology of \emph{diffusion geometry} to model the space of particles.
Namely, we model the set of particles as a sample of a manifold and use distances on this manifold to measure particle similarity.
The diffusion geometry is defined with respect to this manifold and projects each particle into a high-dimensional space, such that Euclidean distances correspond to
robust distances on the manifold, insensitive to imperfections in particle sampling. More specifically, these distances are only low for a pair of particles if there exists a large set
of paths on the manifold between the particles. The distances are parameterized by a \emph{scale} parameter, which only permits paths whose lengths are less than the scale. This allows us to
measure distances in a multi-scale manner.

% uses
The multi-scale property of diffusion geometry allows us to construct continuous variants of separation and clustering of particles.
Particle separation is computed by analyzing
the covariance of the diffusion geometry in a particle's local Euclidean neighborhood. We show how this generalizes the FTLE by allowing the user to inspect
different strengths of separation by adjusting the scale parameter. Grouping is performed in a user-driven manner by selecting individual particles,
and showing neighborhoods determined by the diffusion geometry. For small scales, the neighborhood of a particle are those particles that
remain spatially close over all time steps. For large scales, the neighborhood relates to clusters found in spectral clustering~\cite{ng2001spectral}.
For intermediary scales, the user can inspect the evolution in particle similarity.

% interaction -> teaser
Our interface for visualizing particles uses the separation measure for conveying an overview on features of interest in the data,
in addition to guiding how the user selects particles for inspecting similarity neighborhoods for further details.
Figure~\ref{fig:teaser} shows a simulation of protostar formation from a molecular gas cloud, termed \textsf{Cloud Collapse}. Figure~\ref{fig:teaser}(a)
highlights particle attraction, computed by measuring backward-time particle separation shown at the end of the simulation.
We allow the user to select
a particle slightly offset from an attracting particle, and then visualize its neighborhood. In Figure~\ref{fig:teaser}(b) we select a set of such particles
and their corresponding neighborhoods, each uniquely colored. Hence we can inspect the clustering of particles that are collapsing to the protostar through different
trajectories, as shown in the inset.

Our main contributions are summarized as follows:
\begin{itemize}
	\item We introduce a technique to compute multi-scale similarity between particles in time-varying flow.
	\item We utilize this similarity to construct continuous separation and clustering measures as well as a means to interact with particle
		data guided by these measures.
	\item We show the benefits of our technique for analyzing particle data produced from fluid simulations and astrophysics.
\end{itemize}

\section{Related Work}

% steady flow topology
Our approach is related to works in time-varying vector field visualization that aim to compute separation structures, or
regions where flow is either attracting or repelling. These structures provide counterparts to topological decomposition in steady
flows~\cite{HelmanH90,laramee2007topology}, where a domain partitioning is sought such that each region is comprised of uniform flow.
However in unsteady flow, separation alone does not necessarily yield a full domain decomposition. Certain approaches have dealt with this
by computing vector field topology at each time step and tracking critical points over time~\cite{garth2004tracking}, or considered
improved frames of reference for tracking~\cite{FuchsKSWSHP10}. However, studying time-varying instantaneous topology may poorly reflect
the temporal dynamics of flow.

% LCS
A separation structure frequently used to capture the temporal evolution of flow separation is the Lagrangian Coherent Structure
(LCS)~\cite{haller2001distinguished,shadden2005definition}. The LCS is often computed by detecting ridges in the Finite Time
Lyapunov Exponent (FTLE)~\cite{shadden2005definition}, which measures how much a particle diverges from an infinitesimally-close
neighborhood of particles over a given time interval. Various techniques have been employed to visualize the FTLE and LCS in
2D~\cite{machado2016space} and 3D~\cite{garth2007efficient,sadlo2011time}. Computing the FTLE for 3D flow
data can be expensive, and significant effort has been devoted to efficient techniques that do not
sacrifice approximation quality~\cite{garth2007efficient,sadlo2007efficient}, most relying on a meshing of the domain
to facilitate access of the flow map. Agranovsky et al.~provide efficient computation
of the LCS in a mesh-free manner via moving least squares~\cite{agranovsky2011extracting}. However, Agranovsky et al. assume they can adaptively seed pathlines for vector field integration to interpolate the flow map at arbitrary positions. We do not assume such access, and instead aim to capture admissible details at multiple scales given the particle sampling.

% clustering
One disadvantage with separating structures such as the LCS is that it can overly summarize the data, and additional
detail on flow features may be hard to access. In particular, the LCS does not capture the grouping of flow trajectories, where
a group of flows is characterized by some form of flow similarity. For steady flow, streamline clustering has proven
an effective means of grouping, and there have been many efforts to segment streamlines based on different types of
similarity measures and clustering techniques~\cite{RosslT12,LuCLSW13,ChaudhuriLSW14}. For instance,
R\"ossl and Theisel consider the Haussdorf distance between streamlines as their similarity measure, and establish
a relationship to steady vector field topology~\cite{RosslT12}. Streamline similarity measures have also been used for querying patterns
in flows~\cite{wang2014pattern,tao2016vocabulary} as well as streamline seeding for effective visualization of flow
trajectories~\cite{chen2007similarity,mcloughlin2013similarity}. Clustering and similarity-driven analysis have been extended
to unsteady flow, using generative models~\cite{hong2014flda} or Lagrangian-averaged attributes for dimensionality
reduction~\cite{guo2014scalable}. However, the time-varying scenario presents difficulties due to the temporal
evolution of similarity, wherein clusters can become harder to define. In contrast, our multi-scale similarity enables the user
to explore flow clusters at different levels of detail.

% particles
In response to the growing popularity of particle-based flow simulation, flow visualization of particle data has received recent attention. Agranovsky et al. extract
a collection of particles \emph{in situ}, and then allow the user to extract particles at arbitrary seed positions
\emph{post hoc} through barycentric interpolation~\cite{agranovsky2014improved}. Chandler et al. improved on the interpolation by using
SPH kernels for scattered data interpolation~\cite{ChandlerOJ15}, while follow up
work~\cite{bujack2015lagrangian} considered parametric representations of particles for visualization. Closely related
to our approach is recent work in defining an analog to the FTLE for particle data.
Kuhn et al. construct a triangulation on particle data, and as a discrete representation of time lines they track
the evolution of edges over time, wherein high FTLE values are those edges which tend to collapse in the evolution~\cite{kuhn2014time}.
Shi et al. compute the Jacobian of the flow map directly on particles by linearizing the Jacobian, and subsequently solve for
it in a local neighborhood of particles in a least squares sense~\cite{shi2017analysis}. Our covariance-based measure for particles is related to these approaches, and
generalizes them to the analysis of separation at multiple scales through diffusion geometry.

% coherent sets
Our approach is also inspired by the set of recent works that construct diffusion-type operators on flow
data~\cite{Froyland15,Hadjighasem16,froyland2016dynamic,karrasch2016geometric,banisch2017understanding}.
These approaches are based on the notion of \emph{coherent sets}~\cite{Froyland15}, or spatial regions that are robust to small perturbations
applied to the composition of the forward and backward flow maps. To model the perturbation as a diffusion process
defined on particle flows, these approaches construct different types of diffusion objects such as the Laplace operator~\cite{froyland2016dynamic},
the heat kernel~\cite{karrasch2016geometric}, and in particular space-time diffusion maps~\cite{banisch2017understanding}.
These works mostly consider how to use these objects for flow clustering flow. In contrast, we show
how using diffusion across all scales can provide a more complete picture of flow behavior in particles.

% scale space flow
Our approach employs diffusion geometry, which relies on a notion of scale in constructing similarities between particle trajectories.
This is related to scale-space approaches~\cite{lindeberg2013scale} which analyze field-based data in multiple scales, typically to denoise or find optimal
spatial scales for filtering. Flow analysis has employed scale-space techniques in various contexts, such as vortex
tracking~\cite{bauer2002vortex} and detection of FTLE ridges~\cite{fuchs2012scale}.
However, the construction of multi-scale distances requires different mathematical tools compared to traditional scale-space methods on fields.
Furthermore, it is nontrivial to extend these techniques to particle data, as they utilize a vector field for their
respective scale-space approaches.

% optional: similarity in scivis

\section{Diffusion Geometry of Particles}

Our approach to analyze particle flows uses
the methodology of diffusion geometry~\cite{coifman2006diffusion} to construct similarities between particles. In this section we show how to
construct diffusion geometry on particles, with careful consideration on computing diffusion geometry efficiently and handling spatial locality and nonuniformity for
diffusion.

We assume that we are provided a set $P$ of $n$ particles with starting time $t_1$ over a temporal interval $\tau$, discretized over $T$ time steps $\{t_1, t_2, \ldots, t_T\}$ such
that $t_T = t_1+\tau$. Particle positions are embedded in a $d$-dimensional space where we consider $d=2,3$. For a particle $\mb{p} \in P$,
we denote $\mb{p}_k$ as the particle's position at the time step $t_k$. We refer to a particle's sequence of positions over time as its \emph{trajectory}.
We also assume an arbitrary ordering of the particles in $P$, and denote the $i$'th particle as $\mb{p}^{i} \in P$.

\subsection{Computing Diffusion Geometry}

In the construction of particle similarities we consider the geometric structure that is formed by the set of particles. Namely, we assume
that the set of particles $P$ are sampled from a manifold, called the \emph{particle manifold}, and we measure distances between particles with respect to this manifold.
This perspective is in contrast to techniques that construct features on particles and perform analysis with respect to Euclidean distance in this feature
space~\cite{hong2014flda,wang2014pattern,guo2014scalable}. These feature-based distances are most descriptive for particles that are close in distance, but offer less analytic power for particles that are far.
Our particle-based geometric representation captures meaningful distances on the particle manifold that are both small and large. 

One popular approach for computing distances on manifolds is through geodesic distances, defined as the length of the shortest path
between a pair of points. In practice, geodesics on a sampled manifold are approximated by first forming a graph where an edge exists between a pair of samples if they are in small distance with respect
to some measure, typically Euclidean distance in their embedding space. Geodesics are then found through shortest paths on this graph. A drawback of this technique is that it is
sensitive to incorrect edge connections. For instance, if an edge exists between a pair of samples that are at a large geodesic distance, then all paths
through this edge will be incorrect. This concern is especially problematic for particle data, where the graph construction must face the challenges
of sparse and nonuniform particle sampling.

We adopt a more robust means of computing distances on manifolds, using the framework of \emph{diffusion geometry}~\cite{coifman2006diffusion}. Rather than compute
a single path, diffusion geometry averages the lengths of a set of paths between a pair of samples, where the maximum length of paths is limited by a scale parameter.
Diffusion distances have proven useful in shape analysis~\cite{de2008hierarchical,bronstein2010gromov} compared to geodesics alone, for instance in
handling topological noise due to poor estimation of the local geometry. To extend diffusion geometry to particle data, we perform the following three steps:
\begin{enumerate}
	\item Approximate the local geometry of a particle.
	\item Convert local geometry into a Markov matrix that measures the probability of reaching particles on the manifold.
	\item Compute diffusion geometry based on eigenanalysis of the transition matrix.
\end{enumerate}

\subsubsection{Local Geometry of Particles}

The local geometry of a particle is captured by a similarity kernel $k$, where $k(\mb{p}^i,\mb{p}^j) \in [0,1]$. The kernel only assigns high
similarity to pairs of particles if they are close based on a distance function between particles $d_E$, and a bandwidth parameter $\sigma$ that
limits the spatial extent of $d_E$. We employ a Gaussian kernel for $k$, as is commonly done in manifold learning~\cite{coifman2006diffusion}:
\begin{equation}
	k(\mb{p}^i,\mb{p}^j) = \exp \left(\frac{-d_E^2(\mb{p}^i,\mb{p}^j)}{\sigma^2}\right),
	\label{eq:kernel}
\end{equation}
For the distance function $d_E$ we use the so-called \emph{dynamic distance} between particles~\cite{Hadjighasem16}.  Dynamic distance computes
the time integration of distances between two particles' positions, and it may be discretized as follows:
\begin{equation}
	d_E(\mb{p}^i,\mb{p}^j) = \frac{1}{t_T - t_1} \sum_{k=1}^{T-1} \frac{t_{k+1}-t_k}{2} \left( \lVert \mb{p}^i_{k+1} - \mb{p}^j_{k+1} \rVert + \lVert \mb{p}^i_{k} - \mb{p}^j_{k} \lVert \right)
	\label{eq:dynamicdistance}
\end{equation}
Thus given a particle $\mb{p}^i$, $k$ will report large values to other particles whose distance $d_E$ is small, relative to the bandwidth $\sigma$.
The set of particles with large $k$ captures the local neighborhood of the particle manifold at $\mb{p}^i$.

\subsubsection{Particle Markov Matrix}
\label{sec:markov}

We next construct a Markov matrix based on $k$, denoted $W$, where $W_{ij}$ describes the probability of transitioning to particle $\mb{p}^j$
given particle $\mb{p}^i$ under a random walk on the particle manifold. The Markov matrix is computed by first forming the matrix kernel $K$ where
$K_{ij} = k(\mb{p}^i,\mb{p}^j)$. The $i$'th row of $K$ encodes the local geometry of particle $\mb{p}^i$, and most particles that are not in $i$'s neighborhood will be very small. For computational and memory
efficiency we represent $K$ as a sparse matrix by setting its entries to $0$ if they are less than a threshold. We use $10^{-6}$ in practice.

The matrix $W$ is computed by row normalizing $K$:
\begin{equation}
	W = D^{-1} K,
\end{equation}
where $D$ is a diagonal matrix such that $D_{ii}$ is the summation of all entries in the $i$'th row/column of $K$. Powers of $W$ can be interpreted
as walking a Markov chain forward, in effect diffusing probabilities across the manifold as defined by the local geometry $k$.
$W^s_{ij}$ is the probability of reaching $\mb{p}^j$ from $\mb{p}^i$ for a given exponent $s$, the \emph{scale} parameter. The parameter $s$ may be interpreted
as the \emph{spatial} scale, intrinsic to the manifold, that a random walk may traverse. Large values of $s$ permit large path traversals, thus
increasing the transition probability for particles far apart on the manifold.

\subsubsection{Diffusion Distances of Particles}

Directly using the Markov matrix as our form of particle similarity is problematic due to its asymmetry. Diffusion distances instead consider the overlap in probabilities
over all particles, parameterized by scale. Intuitively, if a pair of particles have mutually common high transition probabilities to other particles, then their diffusion distance
should be low. More specifically, if we denote $(W^s)_{i \cdot}$ as the $i$'th row vector of $W^s$, then the \emph{diffusion distance}, $d_s$, is
\begin{equation}
	d_s(\mb{p}^i,\mb{p}^j) = \lVert (W^s)_i - (W^s)_j \rVert.
\end{equation}
A practical means of computing diffusion distances is through the eigenanalysis of $W$. More specifically, let $\mb{u}^l$, $\lambda_l$ be the $l$'th pair of right eigenvectors
of $W$. Then the diffusion distance can be efficiently computed as
\begin{equation}
	d_s(\mb{p}^i,\mb{p}^j) = \left( \sum_{l \ge 1} \lambda_l^{2 s} (\mb{u}^l_i - \mb{u}^l_j)^2 \right)^{\frac{1}{2}},
	\label{eq:scale}
\end{equation}
where particle $i$ is indexed in $\mb{u}^l$ through subscript. We may also associate an embedding with each particle $\Phi_s(\mb{p}^i) = (\lambda_1 \mb{u}^1_i,\lambda_2 \mb{u}^2_i,...,\lambda_m \mb{u}^m_i)$,
such that the diffusion distance can now be expressed as $d_s^2(\mb{p}^i,\mb{p}^j) = \lVert \Phi_s(\mb{p}^i) - \Phi_s(\mb{p}^j) \rVert^2$.
The embedding $\Phi_s$ is known as the \emph{diffusion geometry}.

Since we would like to faithfully represent the particle manifold through the diffusion distances, the distances should be insensitive to nonuniform sampling of the particles.
To handle this we renormalize the entries of $W$ in order to factor out density. Renormalizing has the effect of recovering the
intrinsic geometry of the manifold, see Coifman and Lafon~\cite{coifman2006diffusion} for details.

\subsection{Landmark Diffusion Distances}
\label{sec:landmark}

% motivation
A practical issue with computing diffusion distances is the computational complexity of performing the eigendecomposition of the matrix $W$.
Although $W$ is sparse, as discussed in Section~\ref{sec:markov}, its eigendecomposition can still be expensive when the number of particles $n$ becomes large,
for instance exceeding $10^6$.

% approach - relationship to diffusion distances
To address this we employ a landmark technique to approximate the eigenvectors and eigenvalues of $W$. We extend the landmark-based spectral clustering technique of Chen \& Cai~\cite{chen2011large} to
the case of diffusion geometry. Namely, suppose that we have a set of $n_l$ landmarks $I \subset \{1 ... n\}$. We compute affinities $k$ between each landmark particle $\mb{p}^i$, $i \in I$ and every other
particle $\mb{p}^j$ to form a landmark similarity matrix $K_l \in \mathbb{R}^{n \times n_l}$. We then use $K_l$ to form a low-rank factorization of $W^2$ by constructing $W_l = D^{-1}_r K_l D^{-1}_c (K_l D^{-1}_c))^{\intercal}$. Here $D_r \in \mathbb{R}^{n \times n}$ and $D_c \in \mathbb{R}^{n_l \times n_l}$ are diagonal matrices formed by taking the row sums and column sums of $K_l$, respectively, in order to yield the normalized affinity matrix $W_l$.
Since $W_l$ is of rank at most $n_l$, its eigendecomposition can be computed from a sparse matrix of dimension $n_l \times n_l$~\cite{chen2011large}. Assuming $W_l$ is
a good approximation to $W^2$, then its eigenvectors and eigenvalues will also be close, hence our diffusion distances will be accurate estimates.

% landmark selection
The quality of this approximation is dependent on the chosen landmarks. We would like the landmarks to provide a good geometric coverage of the domain, in order
to capture flow features in the data. However, we would also like our landmark scheme to be computationally efficient, and scale well as a function
of the number of particles and the number of time steps.

To this end, we perform a modified version of \emph{farthest point sampling}. First, we select a
particle at random to be a landmark. The next selected particle is the one furthest in distance to the previously chosen particle. We use $d_E$ to measure distance
between particles, but only at uniformly-spaced temporally subsampled positions for efficiency considerations, in practice using a temporal resolution of $5$. Each subsequent
landmark is chosen such that its minimum distance to previously sampled landmarks is the largest amongst candidate particles. We find this scheme produces landmarks
representative of features in the data, and is reasonably efficient to compute. It has complexity $O(n \cdot n_l)$, compared to using all time steps, which would be $O(n \cdot n_l \cdot T)$.
Nevertheless, when both $n$ and $n_l$ are large then landmark selection can become the computational bottleneck. We conduct an experimental evaluation on performance and computational efficiency in Section~\ref{sec:landeval}.

\subsection{Bandwidth Selection}
\label{sec:bandwidth}

% motivation
The spatial bandwidth $\sigma$ (Eq.~\ref{eq:kernel}) affects the interpretation of the diffusion distances. We would like $\sigma$ to be in the proper range to capture the local geometry of the manifold $k$.
Particle paths, however, present two main difficulties in using a global $\sigma$ for similarity. First, the scale of particle positions as a function of time can significantly change, for instance
in astrophysics simulations particles are often expanding or collapsing over time. Secondly, within a time step the density of particles can vary across different spatial regions, and a single $\sigma$
may fail to capture the geometric structure of all regions.
%Note that variability in density is different from nonuniform sampling.

% space-time bandwidths, designed for landmarks
To handle these cases for our landmark-based formulation, we employ spatiotemporal bandwidths $\sigma_{k,i}$ for each time step $t_k$ and particle landmark $i$. More specifically, for particle $\mb{p}^i$ at time $t_k$ we gather
its nearest neighbors, where we fix the number of neighbors to 6, and take $\sigma_{k,i}$ as the average distance to its nearest neighbors. Then for an arbitrary particle $j$, its kernel similarity to particle landmark $i$ is modified as follows:
\begin{equation}
	k(\mb{p}^i,\mb{p}^j) = \exp \left(-\frac{1}{\tau} \sum_{k=1}^{T} \frac{\lVert \mb{p}^i_{t_k} - \mb{p}^j_{t_k} \rVert^2}{(\alpha \sigma_{k,i})^2} \right),
	\label{eq:bandwidth}
\end{equation}
where without loss of generality we assume uniformly-spaced temporal sampling.
$\alpha$ is a user-defined parameter that scales the bandwidth, and we experimentally found $\alpha \in (0.75,1.75)$ to be reasonable.
The expression in the exponential is a generalization of the distance function $d_E$, taking into account sampling variation across space and time.
Furthermore, note that it is only necessary to define bandwidths with respect to the landmark particles since the distribution of landmarks limits the scale at which we
can analyze the geometry.

\subsection{Particle Separation and Time}

Diffusion geometry of particles is also dependent on the time parameters $t$, $\tau$, and in particular $T$. For $T=2$ a particle is comprised of its starting and ending position.
In this case $k$ will only report high similarity for particles that start and end close to each other. This behavior is related
to the FTLE, which measures the maximal spatial distortion of the flow map $\Psi_t^{\tau} : \mathbb{R}^d \rightarrow \mathbb{R}^d$ defined from $t$ to $t+\tau$. More specifically,
distortion in the flow map correlates with high variation in $k$ for particles close in spatial position at $t_1$.

In contrast, for $T > 2$ particles must remain close over all times for $k$ to report high similarity.
The appropriate choice of time discretization is application dependent. Analysis with $T=2$ does not consider particle positions at intermediary times, whereas for $T>2$
the averaging of particle positions can make particle separation less discriminative.
Unless otherwise specified we analyze particles using all available time samples. 
Where comparison is appropriate we use a time-averaged analogue of the traditional FTLE, which we denote t-FTLE.  Specifically, we replace $\Psi$ with $\bar{\Psi}$ a modified flow map whose
range is the space spanned over all particles in the time range $\{t_2, t_3, \ldots, t_T\}$, i.e. $\bar{\Psi}_t^{\tau} : \mathbb{R}^d \rightarrow \mathbb{R}^{(T-1)d}$.
The maximal spatial distortion of $\bar{\Psi}$ measures average particle distortion over all sampled times.

\subsection{Diffusion Scale}

The scale $s$ in Equation~\ref{eq:scale} that parameterizes diffusion distances enables a meaningful and robust computation of particle separation and similarity.
Diffusion distances for a pair of particles will be small only if these particles are connected by paths on the manifold whose lengths are proportional to the scale.
We find that particles separated by LCS have high diffusion distance since manifold paths must travel around the ridge. Thus, the scale acts as an indicator of separation strength; specifically, it is proportional to the distance on the manifold necessary to cross LCS.
Conversely, if the flow has weak separation, such as mostly laminar flow and swirling motions, then local similarity as captured by $k$ in Equation~\ref{eq:kernel} will rapidly
diffuse on the manifold for small scales. Particles in such regions will have low diffusion distances at small scales, indicating that they are highly similar.

Furthermore, diffusion distances are computed using a set of paths on the manifold, not just a single path. Thus diffusion distances are robust to small
errors in local geometry approximation, particularly as the scale increases~\cite{bronstein2010gromov}. For particle separation,
if a small set of particle pairs separated by a ridge are similar according to Equation~\ref{eq:kernel}, then diffusion distances are capable
of filtering out these false positives. For particle similarity, false negatives defined by particles not similar based on Equation~\ref{eq:kernel}
will nevertheless have low diffusion distance if a sufficiently large number of nearby particles are highly connected on the manifold.

We make these concepts more precise through an example, using an analytically-defined flow map that has a separation ridge which varies in strength.
We extend the \textsf{Sine Ridge} example of Kuhn et al.~\cite{kuhn2012benchmark}, whose flow map is characterized by a sine-shaped ridge of a user-specified
strength. To vary separation strength we use Kuhn et al.'s domain deformation~\cite{kuhn2012benchmark} and modify the flow map as follows:
\begin{equation}
	\phi_{\textsf{sine}}(\mb{x},t) = \begin{pmatrix} \frac{x}{\sqrt{\mb{x}^2 + (1-\mb{x}^2) e^{-2 t p(y)}}}\\
		y+t \end{pmatrix}
\end{equation}
where $t$ is the time step and $p$ is a function that controls the ridge strength. The function $p$ is defined such that $p(-4) = 0.05$ and $p(4) = 4$
given the spatial domain $[-2,2] \times [-4,4]$, and performs linear interpolation as a function of the $y$ coordinate. High values of $p$ result in the
divergence of particles seeded near the center of the sine function. We perform uniform seeding in the spatial domain
and consider particles starting at $t_1 = 0$ for a duration of $\tau = 1$, for $T = 100$ time steps.

\begin{figure}[!t]
	\begin{tikzpicture}
		\node[inner sep=0pt,anchor=south west]  at (0,0)
		{\includegraphics[width=1.0\linewidth]{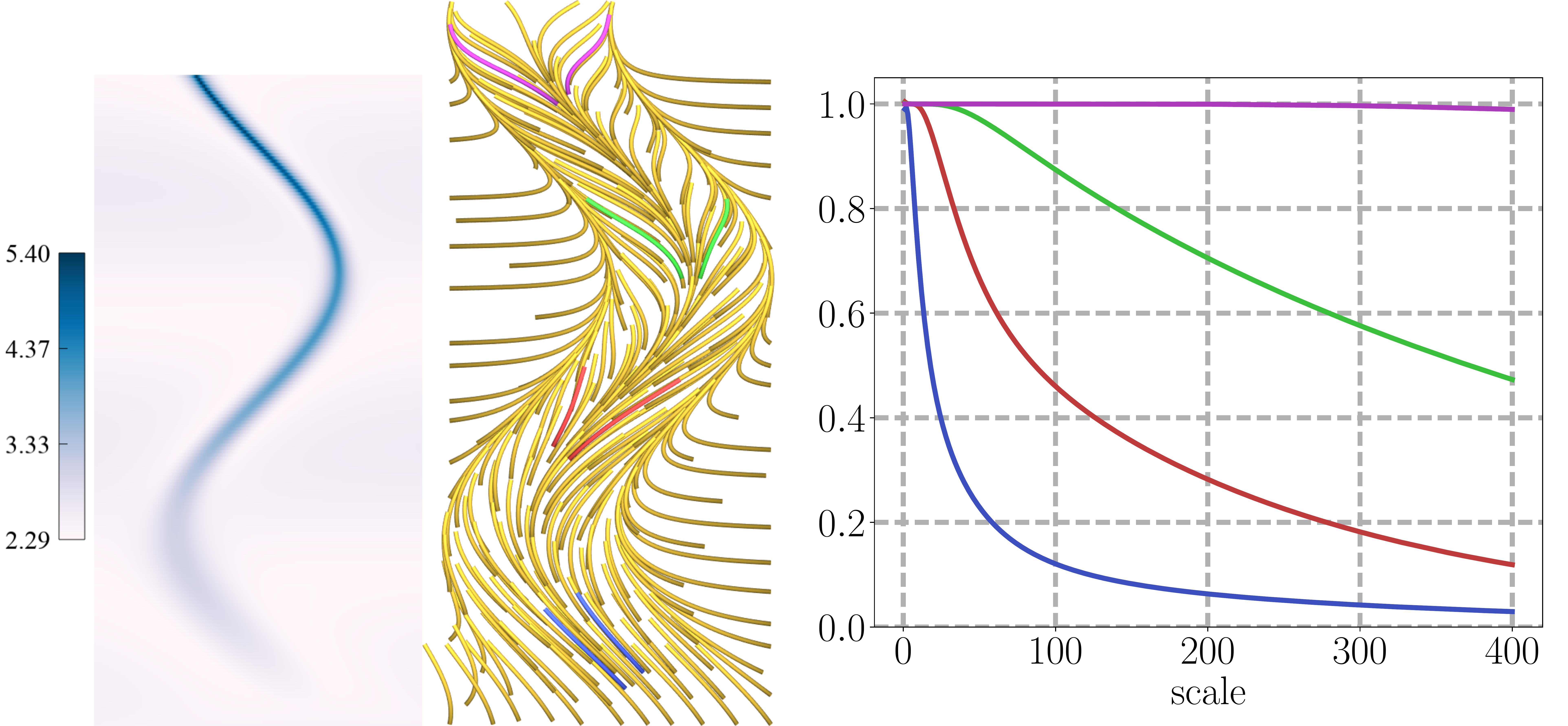}};
		\node at (1.5,-0.25) {(a) t-FTLE};
		\node at (3.6,-0.25) {(b) Particles};
		\node at (7.0,-0.25) {(c) Diffusion Distances};
	\end{tikzpicture}
	\caption{We show the t-FTLE (a) and sample particles (b) for \textsf{Sine Ridge}, and the diffusion distances (c) as a function of scale,
	where the colors of each curve (c) correspond to the colored pairs of particles (b). Note that pairs of particles that cross
	stronger regions of the ridge have a higher diffusion distance across scale than those that cross weaker regions.  Sample particles increase in brightness to indicate forward time.}
	\label{fig:sineridge}
\end{figure}
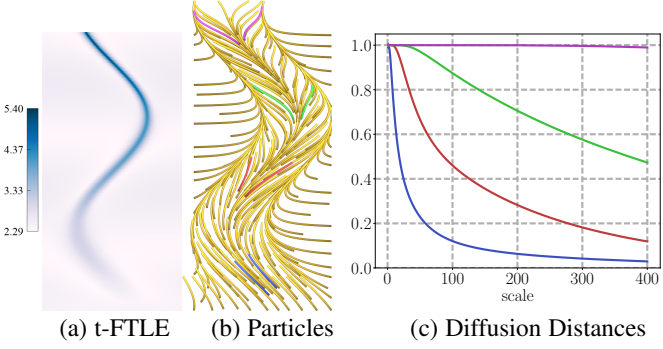

In Figure~\ref{fig:sineridge}(a) we show the t-FTLE, whose strength increases from the bottom to the top of the domain due to the ridge strength function $p$.
We show a subset of example trajectories in Figure~\ref{fig:sineridge}(b) that demonstrate this divergence, and highlight four pairs of particles that cross the ridge.
We compute diffusion distances over a sequence of scales for these particles as shown in Figure~\ref{fig:sineridge}(c). To compare distances across scale
we divide the squared distances by their squared norms. Namely for particles $\mb{p}^i$ and $\mb{p}^j$ we divide $d_s(\mb{p}^i,\mb{p}^j)$ by
$(\lVert \Phi_s(\mb{p}^i) \rVert^2 + \lVert \Phi_s(\mb{p}^j) \rVert^2)$, a standard technique for normalizing diffusion geometry~\cite{sun2009concise}.

We observe that particle pairs on opposite sides of the ridge where the strength is low result in diffusion distances that sharply decrease as a function of scale, and
vice versa for particle pairs where the ridge strength is high. Since particles that cross ridges are unlikely to be connected in their local geometry,
the spatial scale necessary for paths to connect these particles on the manifold is large, and is a function of the shape of the ridge. Thus
the red-colored pair's diffusion distance is smaller than the green pair's, due to a smaller spatial scale necessary to cross the ridge.

\begin{figure}[t]
	\begin{center}
	\begin{subfigure}[!t]{0.2\textwidth}
		\includegraphics[width=\linewidth]{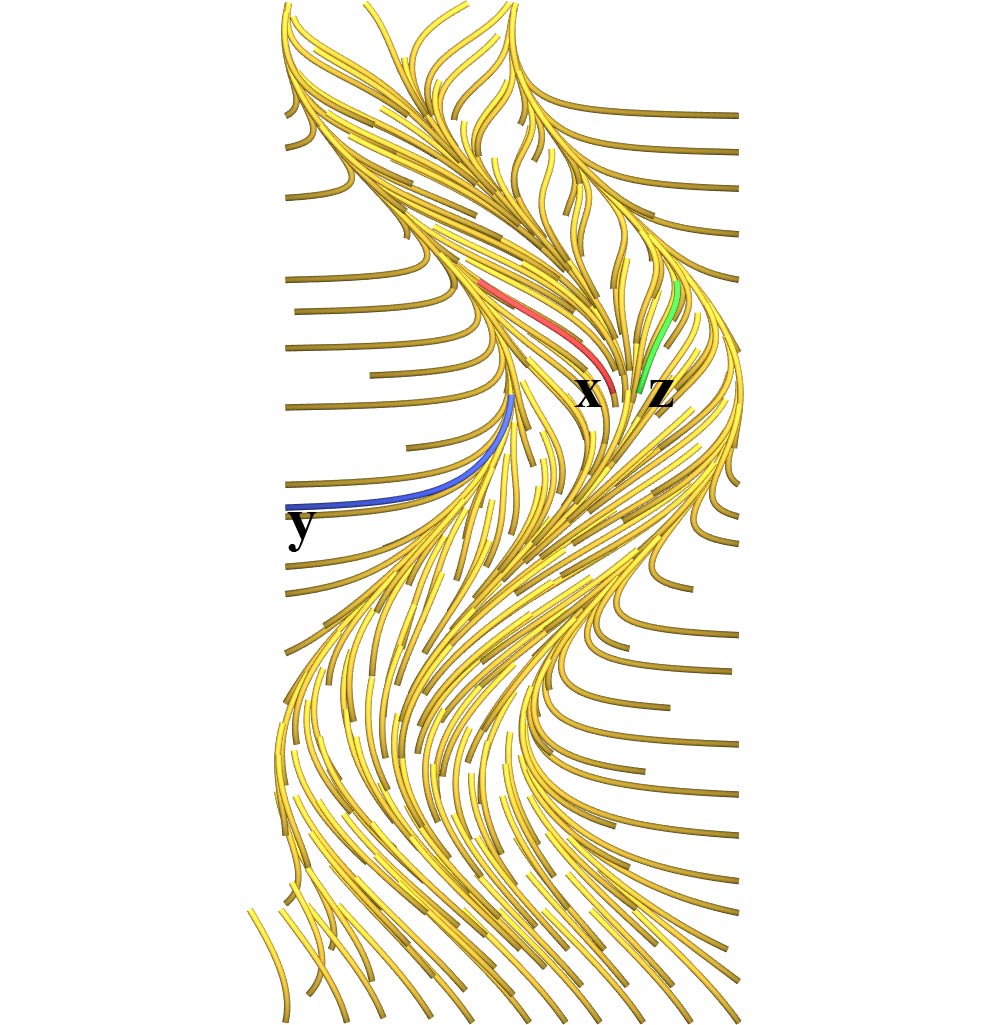}
		\caption{Selected particles}
	\end{subfigure}
	\begin{subfigure}[!t]{0.24\textwidth}
		\includegraphics[width=\linewidth]{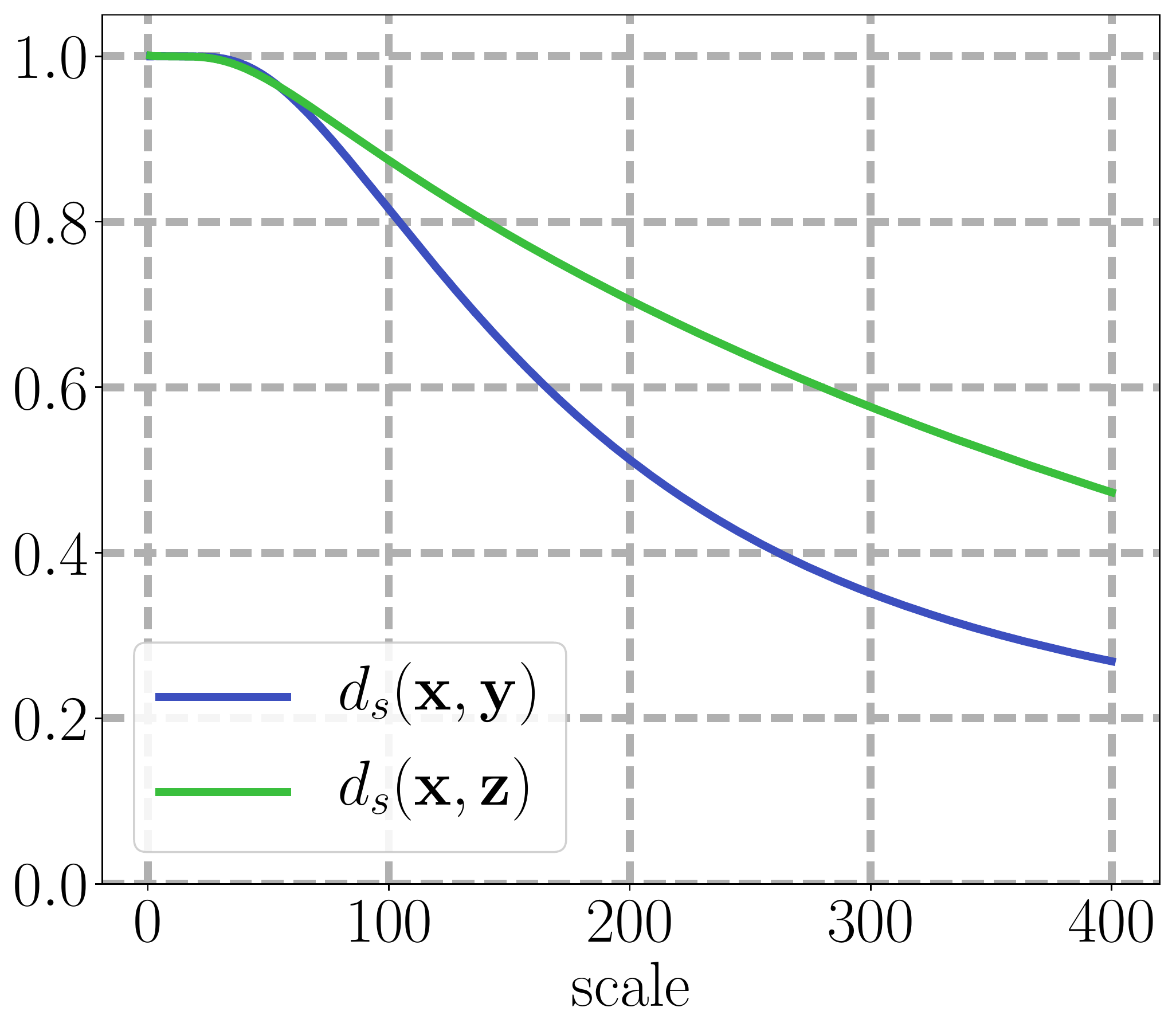}
		\caption{Diffusion distances}
	\end{subfigure}
	\begin{subfigure}[!t]{0.13\textwidth}
		\includegraphics[width=\linewidth]{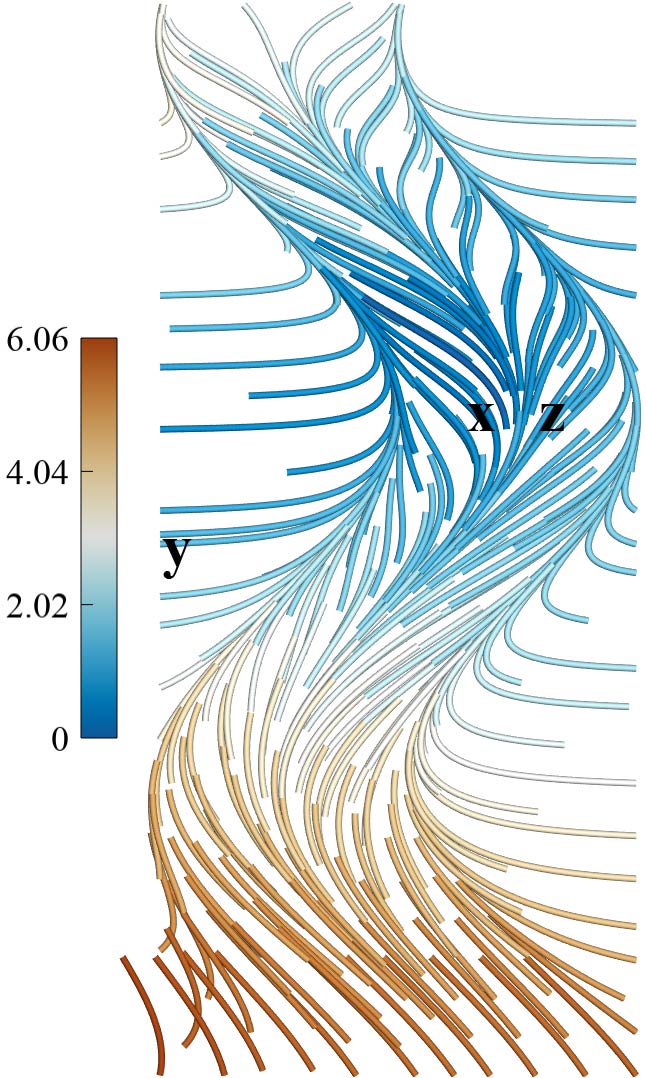}
		\caption{$d_E(\mb{x},\cdot)$}
	\end{subfigure}
	\begin{subfigure}[!t]{0.13\textwidth}
		\includegraphics[width=\linewidth]{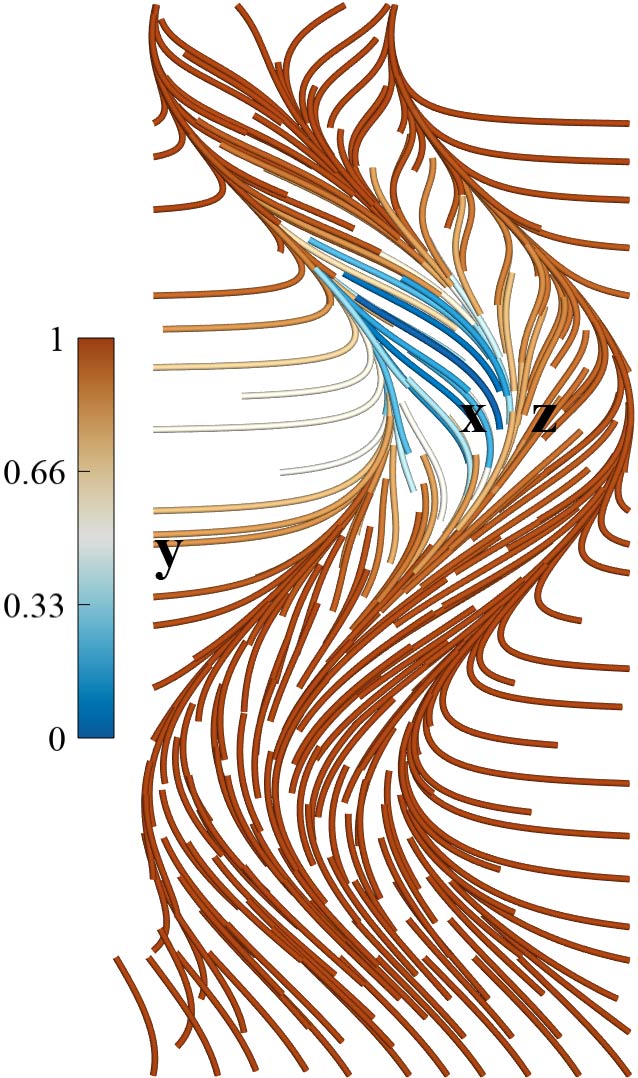}
		\caption{$d_{100}(\mb{x},\cdot)$}
	\end{subfigure}
	\begin{subfigure}[!t]{0.13\textwidth}
		\includegraphics[width=\linewidth]{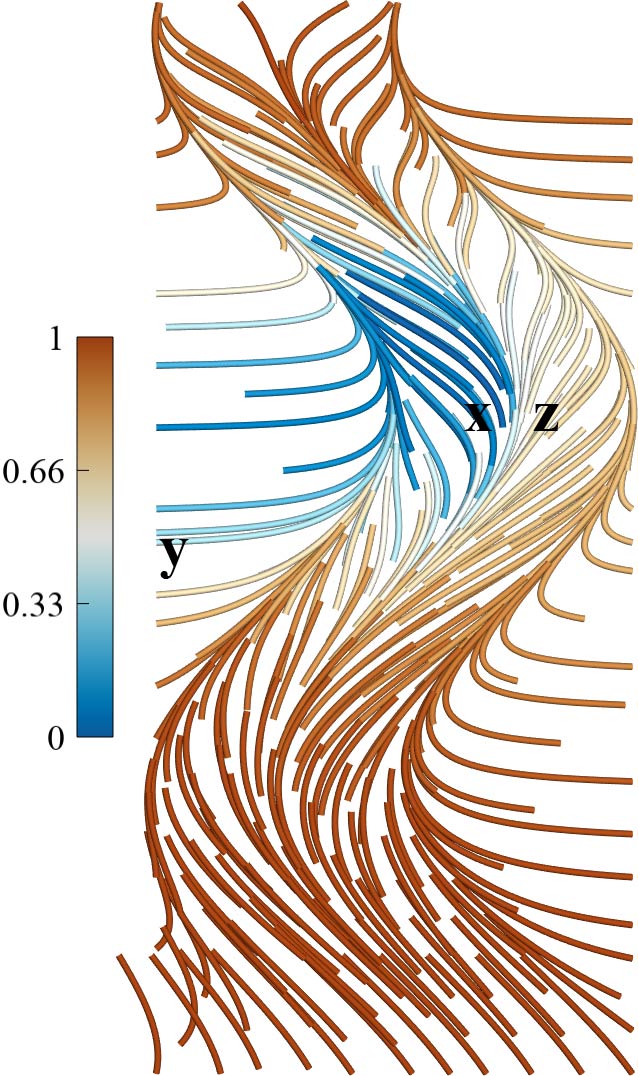}
		\caption{$d_{300}(\mb{x},\cdot)$}
	\end{subfigure}
	\end{center}
	\caption{We highlight selected particles $\mb{x}$, $\mb{y}$, and $\mb{z}$ in \textsf{Sine Ridge} (a) and compare $d_s(\mb{x},\mb{y})$ to $d_s(\mb{x},\mb{z})$.
	Distances from $\mb{x}$ as defined by $d_E$ are color-mapped (c) compared to $d_s$ in (d) and (e) for $s = 100$ and $s = 300$, respectively. Note
	that $d_E(\mb{x},\mb{z}) < d_E(\mb{x},\mb{y})$ despite the separation ridge crossing $\mb{x}$ and $\mb{z}$. Diffusion distances
	respect separation, as shown in (b) and (d)-(e) as we increase scale.}
	\label{fig:sinecompare}
\end{figure}

We next compare diffusion distances for a pair of particles that cross a ridge, and a pair of particles that do not. Figure~\ref{fig:sinecompare}(a)
highlights three particles in the \textsf{Sine Ridge} flow, and we compare the distances $d_s(\mb{x},\mb{y})$ to $d_s(\mb{x},\mb{z})$ for a
sequence of scales in Figure~\ref{fig:sinecompare}(b). We observe that for small scales both distances are large, but as the scale increases
$d_s(\mb{x},\mb{y})$ becomes smaller than $d_s(\mb{x},\mb{z})$. The separation ridge causes paths on the manifold between $\mb{x}$ and $\mb{y}$ to connect at a smaller
scale than those between $\mb{x}$ and $\mb{z}$.

Figure~\ref{fig:sinecompare}(c) shows particles that are color-mapped based on distances from $\mb{x}$ as defined by $d_E$ in Equation~\ref{eq:dynamicdistance}.
We find that $d_E(\mb{x},\mb{z}) < d_E(\mb{x},\mb{y})$, despite no ridge present between $\mb{x}$ and $\mb{y}$.
In Figures~\ref{fig:sinecompare}(d) and (e) we compare to diffusion distances from $\mb{x}$
for scales $s = 100$ and $s = 300$ respectively. Note that as $s$ increases, $d_s(\mb{x},\mb{y}) < d_s(\mb{x},\mb{z})$. Thus our particle manifold model
enables us to compute distances that respect separation ridges in flow, unlike using more standard distances such as $d_E$. Furthermore,
note that $d_s$ does in fact decrease as the scale increases for particles crossing the ridge. This is because the local geometry estimation of
Equation~\ref{eq:kernel} assigns similarity to particles that cross the ridge. However, $d_s$ remains lower for particles near $\mb{y}$ due
to the much larger number of paths connecting particles on the left side of the ridge.

\section{Computing and Visualizing Particle Separation}

In this section we illustrate how to use diffusion geometry for the purposes of computing and visualizing particle separation.
We use the multi-scale property of our distances to compute a scale-based proxy for the FTLE for particle data.
The user may interactively adjust the scale to visualize how particle strength varies.

\subsection{Multi-Scale Particle Separation}

We first show how to approximate the FTLE and t-FTLE from particle data, and then how this may be extended with diffusion geometry.
Recall that the FTLE analyzes the Jacobian $J$ of the flow map. We are only given a sampling of the flow map in the form of particle data, thus
we are unable to approximate the Jacobian through finite differences of the flow map, as is traditionally done~\cite{haller2000lagrangian,shadden2005definition}.
Instead, we can approximate the Jacobian through the covariance of the particles:
\begin{equation}
	C(\mb{p}^i) = \sum_{\mb{p}^j \in N(\mb{p}^i)} (\mb{p}^j_{T'} - \mb{p}^i_{T'}) (\mb{p}^j_{T'} - \mb{p}^i_{T'})^{\intercal},
	\label{eq:particlecov}
\end{equation}
where $N$ denotes a local neighborhood of particles with respect to a given time step, either $t_1$ or $t_T$, and $\mb{p}^j_{T'}$ denotes the vector of all positions excluding
the originating time, either $[t_2,t_T]$ or $[t_1,t_{T-1}]$, which respectively corresponds to forward-time or backward-time separation.
It can be shown~\cite{singer2008non} that, under certain sampling conditions, the covariance is a first-order approximation to $J^{\intercal} J$,
the right Cauchy-Green tensor, and is the mathematical object used to compute the FTLE~\cite{shadden2005definition}. Namely, we approximate the FTLE as:
\begin{equation}
	\gamma(\mb{p}^i) = \frac{1}{\tau} \log\left(\lambda_1\left(C(\mb{p}^i)\right)\right),
\end{equation}
where $\lambda_1$ is the maximum eigenvalue. Note that for $T=2$ and $T>2$ this is an approximation of the FTLE and the t-FTLE, respectively.

\begin{figure}[t]
	\begin{center}
	\begin{subfigure}[b]{0.22\textwidth}
		\includegraphics[width=\linewidth]{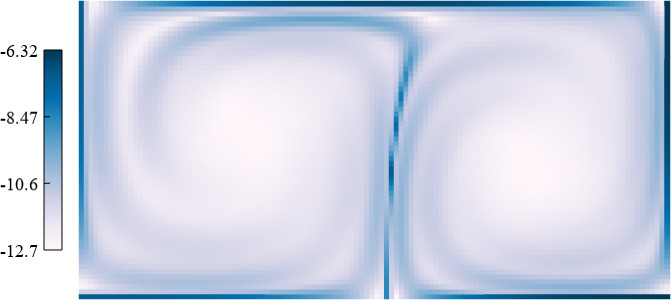}
		\caption{FTLE}
	\end{subfigure}
	\begin{subfigure}[b]{0.22\textwidth}
		\includegraphics[width=\linewidth]{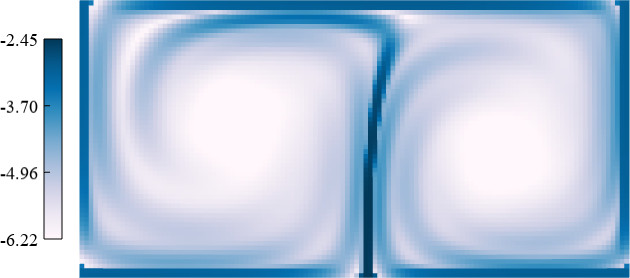}
		\caption{Particle Path Separation}
	\end{subfigure}
	\begin{subfigure}[b]{0.22\textwidth}
		\includegraphics[width=\linewidth]{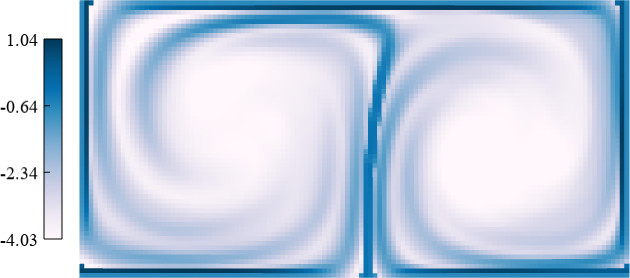}
		\caption{Diffusion Separation $s=28$}
	\end{subfigure}
	\begin{subfigure}[b]{0.22\textwidth}
		\includegraphics[width=\linewidth]{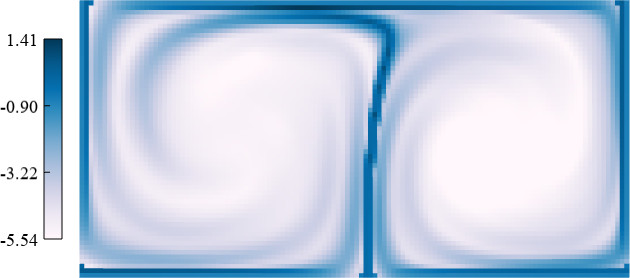}
		\caption{Diffusion Separation $s=78$}
	\end{subfigure}
	\begin{subfigure}[b]{0.22\textwidth}
		\includegraphics[width=\linewidth]{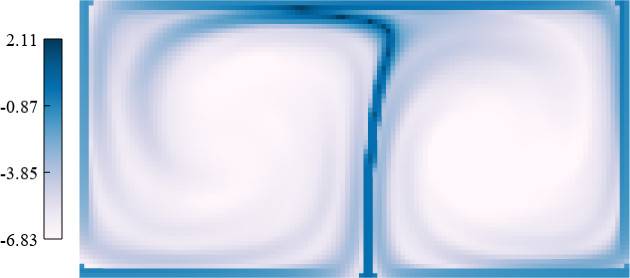}
		\caption{Diffusion Separation $s=145$}
	\end{subfigure}
	\end{center}
	\caption{We show particle and diffusion-based separation measures (b)-(e) compared to FTLE (a) for uniformly-seeded particles. For diffusion separation, increasing the
	scale filters out weaker ridges, leading to similar separation as the FTLE, up to a scale.}
	\label{fig:dg-uniform_separation}
\end{figure}

We can extend this formulation with our multi-scale similarities by directly using the diffusion geometry $\Phi_s$ in place of the particle positions. More
specifically, the diffusion geometry-based covariance, or diffusion covariance, follows as:
\begin{equation}
	C_s(\mb{p}^i) = \sum_{\mb{p}^j \in N(\mb{p}^i)} (\Phi_s(\mb{p}^j) - \Phi_s(\mb{p}^i)) (\Phi_s(\mb{p}^j) - \Phi_s(\mb{p}^i))^{\intercal},
	\label{eq:diffusioncov}
\end{equation}
and the corresponding multi-scale FTLE for particles:
\begin{equation}
	\gamma_s(\mb{p}^i) = \frac{1}{\tau} \log\left(\lambda_1\left(C_s(\mb{p}^i)\right)\right),
\end{equation}
where $s$ is the scale parameter. For small $s$ the diffusion distances are proportional to Euclidean distances in a small neighborhood, up to a scale factor that depends
on the local kernel similarity $k$. Since the eigenvalues of the covariance matrix are invariant to translation and rotation, $\gamma_s$ will be proportional to $\gamma$ in
this setting. As $s$ increases, $\gamma_s$ will only preserve details captured in diffusion distances. For instance, weak ridges will
naturally be filtered out since a large $s$ implies that particles separated by such ridges will eventually become similar, due to the diffusion of similarities.
Analogous with Equation~\ref{eq:particlecov}, if $N$ is defined with respect to $t_1$ then we compute forward-time separation, or particle repulsion. Conversely, if $N$ is defined with
respect to $t_T$ then we compute backward-time separation, or particle attraction. Unless otherwise specified, for all results in the paper we use
forward-time separation.

\textbf{2D Separation Example} To highlight our separation measure, we use the \textsf{Double Gyre} dataset~\cite{shadden2005definition}. It is defined as follows:
\begin{eqnarray*}
v(x,y,t)_x & = & -\pi A\sin(\pi f(x,t))\cos(\pi y) \\
v(x,y,t)_y & = & \pi A\cos(\pi f(x,t))\sin(\pi y) \tfrac{\partial}{\partial x} f(x,t)
\end{eqnarray*}
where
\begin{eqnarray*}
f(x,t) & = & \varepsilon\sin(\omega t)x^2 + (1 - 2\varepsilon\sin(\omega t))x 
\end{eqnarray*}
and $A = 0.1$, $\omega = \frac{\pi}{5}$, and $\varepsilon = 0.1$.
It is characterized by two distinct regions of recurring swirling motion that are separated by a dominant ridge.
We uniformly seed particles in the domain $[0,2] \times [0,1]$, and compute particles by integrating the vector field starting at $t_1 = 0$
for a duration of $\tau = 2\pi$. Since the \textsf{Double Gyre} is a standard benchmark dataset for FTLE computations~\cite{shadden2005definition,fuchs2012scale}, we follow this convention and set $T = 2$.
%The FTLE is a useful quantity for finding this ridge, as it defines the spatial distortion of particles in a local region
%integrated over a fixed time -- high values of the FTLE indicate regions of flow separation.

\begin{figure}[t]
	\begin{center}
	\begin{subfigure}[b]{0.22\textwidth}
		\includegraphics[width=\linewidth]{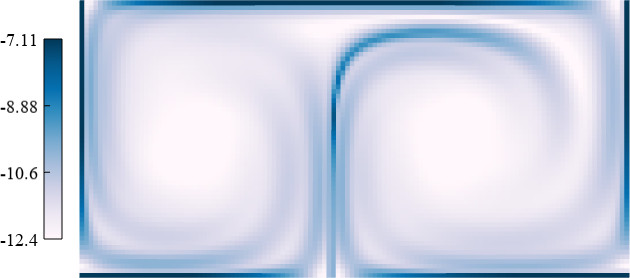}
		\caption{FTLE}
		\label{subfig:dgsep-ftle-nonuniform}
	\end{subfigure}
	\begin{subfigure}[b]{0.22\textwidth}
		\includegraphics[width=\linewidth]{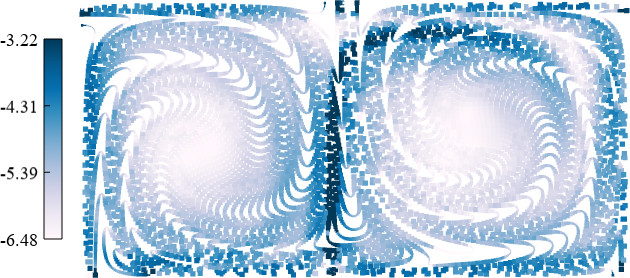}
		\caption{Trajectory Separation}
		\label{subfig:dgsep-covtraj-nonuniform}
	\end{subfigure}
	\begin{subfigure}[b]{0.22\textwidth}
		\includegraphics[width=\linewidth]{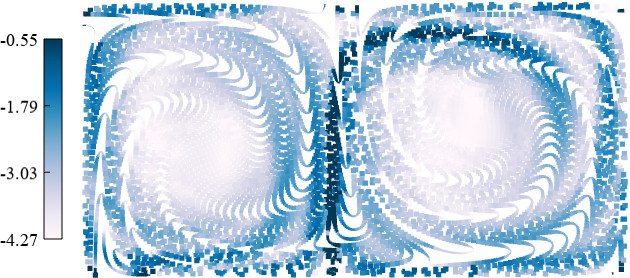}
		\caption{Diffusion Separation $s=28$}
		\label{subfig:dgsep-covtraj-nonuniform}
	\end{subfigure}
	\begin{subfigure}[b]{0.22\textwidth}
		\includegraphics[width=\linewidth]{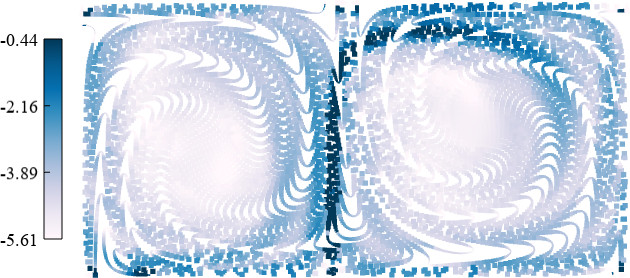}
		\caption{Diffusion Separation $s=78$}
		\label{subfig:dgsep-covtraj-nonuniform}
	\end{subfigure}
	\begin{subfigure}[b]{0.22\textwidth}
		\includegraphics[width=\linewidth]{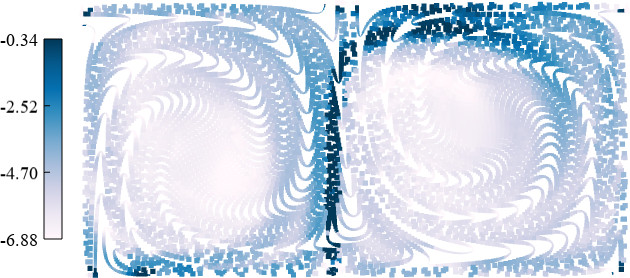}
		\caption{Diffusion Separation $s=145$}
		\label{subfig:dgsep-covtraj-nonuniform}
	\end{subfigure}
	\end{center}
	\caption{We illustrate our separation measures (b)-(e) compared to FTLE (a) when the seeding is nonuniform. We are still able to capture the primary ridge, as well
	as weaker separation regions, despite the nonuniformity.}
	\label{fig:dg-nonuniform_separation}
\end{figure}

Figure~\ref{fig:dg-uniform_separation}(a) shows the FTLE, Figure~\ref{fig:dg-uniform_separation}(b)
shows the particle-based covariance measure $\gamma$, and Figures~\ref{fig:dg-uniform_separation}(c-e) show our diffusion-based covariances
for different scales $\gamma_s$.
We find that our separation measures are able to capture the main ridge in the \textsf{Double Gyre}, however note that scale $s=145$ results in the filtering of weaker
ridges, and better corresponds to the FTLE.

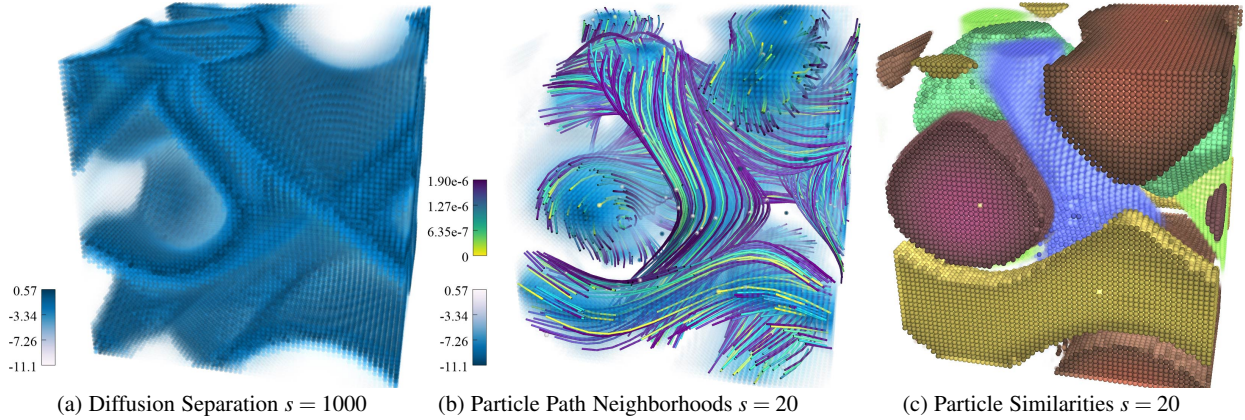
\begin{figure*}[!t]
	\centering
	\begin{tikzpicture}[thick,scale=0.9, every node/.style={transform shape}]
		\node[inner sep=0pt,anchor=south west]  at (0,0)
		{\includegraphics[width=1.0\linewidth]{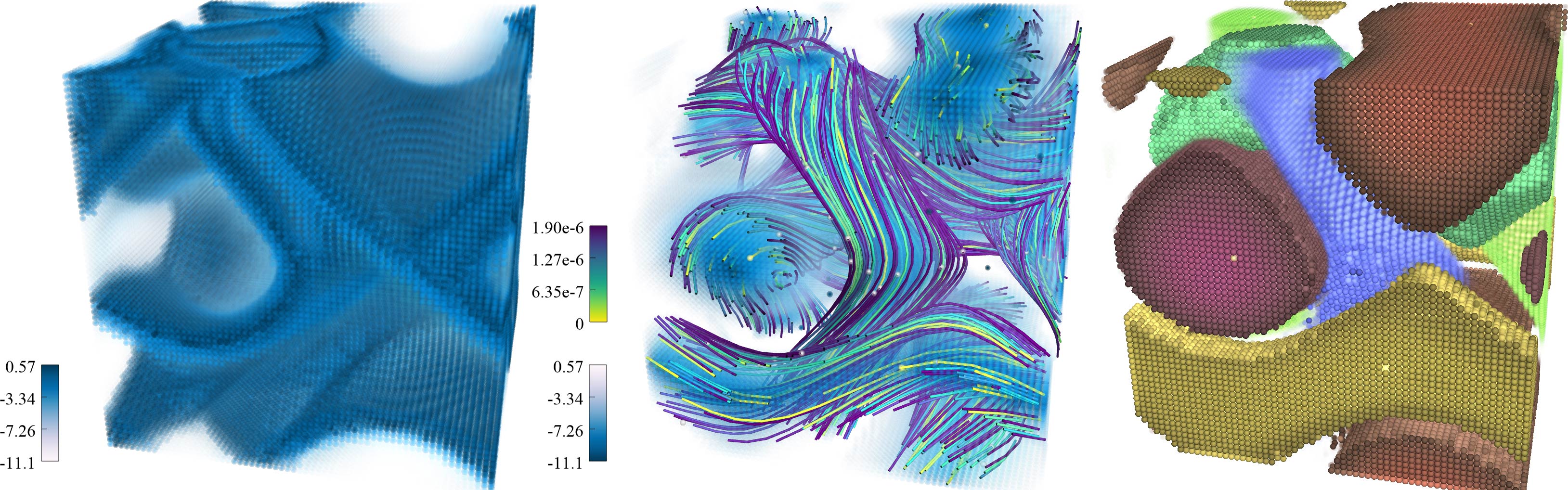}};
		\node at (3,-0.25) {(a) Diffusion Separation $s = 1000$};
		\node at (9,-0.25) {(b) Particle Path Neighborhoods $s = 20$};
		\node at (15.25,-0.25) {(c) Particle Similarities $s = 20$};
	\end{tikzpicture}
	\caption{We show how diffusion-based separation (a) and similarity (b)-(c) aids in extracting the six clusters of the \textsf{ABC} steady flow dataset. The neighborhood
	trajectories (b) depict each cluster from (a) as having a swirling motion, contained by the separating boundaries (a).}
	\label{fig:steadyabc}
\end{figure*}

In Figure~\ref{fig:dg-nonuniform_separation} we show a similar setup, except that we consider particles starting at $t_1 = 2\pi$ for a duration of $\tau = 2\pi$,
wherein particles are uniformly seeded at time $0$ and integrated to time $4 \pi$. In contrast to the previous example, this results in a highly
nonuniform set of particle positions at $t_1$, thus posing an additional challenge for estimating particle separation. We see that our covariance measures are fairly insensitive
to the sampling distribution, and able to capture the primary ridge as well as weaker ridges, with increasing scale filtering these out.

% visualizing separation in 3D: opacity map, go over 3D steady ABC
\textbf{3D Separation Example} We next show our diffusion-based separation measure for an analytic steady periodic 3D flow known as the \textsf{ABC} (Arnold-Beltrami-Childress) Flow.
The vector field equations are defined as follows:
\begin{equation}
	v(x,y)= \begin{pmatrix} A \sin(z) + C \cos(y) \\
	                        B \sin(x) + A \cos(z) \\
									C \sin(y) + B \cos(x) \end{pmatrix}.
\end{equation}
We consider particles that are uniformly-seeded in the domain $[0,2\pi]^3$ and integrate particles starting at $t_1 = 0$ for duration $\tau = 8$, with $T = 40$ samples.
To visualize the separation field in 3D we opacity map particles via the transfer function $\hat{\gamma_s}^{a}$, where $\hat{\gamma_s}$ rescales $\gamma_s$ to between
$[0,1]$ based on the minimum and maximum values of $\gamma_s$ across all particles, and $a$ is a parameter the user can tune to adjust the opacity. Figure~\ref{fig:steadyabc}(a)
shows the diffusion-based separation for the \textsf{ABC} flow. We can clearly make out sheets which partition the domain into regions of uniform flow.

\section{Visualizing Particle Similarity}

% color-mapping and opacity-mapping similarity for particles at starting points
Diffusion distances are used to compute particle similarity, such that a low distance implies high similarity. In this section we discuss techniques
for visualizing particle similarity through interacting with diffusion distances. We first describe techniques for visualizing and interacting
with particle similarity, and then discuss two important characteristics of particle-based diffusion distances that enable meaningful
user interaction: objectivity, and the relationship to coherent sets.

\subsection{Interacting with Diffusion Distances}

We have developed two techniques to visualize particle similarity. First, we allow the user to select a particle, and form a scalar field by computing diffusion distances to
to all other particles. Particles are color-mapped based on this field, and in 3D we opacity map particles in the same manner as separation, allowing the user to adjust transparency.
Furthermore, we enable the user to compare diffusion distances by selecting multiple particles, followed by applying a unique color map to each distance scalar field,
and averaging the fields to produce a single scalar field. The brightness value of each color encodes distance, where small distance
is proportional to high brightness.
%This form of selection is used to visualize distances in Figures~\ref{fig:objectivity} and~\ref{fig:fourcenters}.

% seeding particle trajectories using diffusion distance neighborhoods
We additionally use the diffusion distances for similarity-based seeding in visualizing particle trajectories. Namely, for a given selected particle, we gather all particles
whose diffusion distance is less than a fixed threshold. To reduce clutter, we allow the user to set a maximum number of particle trajectories, and subsample particles through farthest point sampling
under diffusion distance, in order to ensure well-spaced particle particle trajectories with respect to the diffusion geometry. Increasing the scale naturally expands the neighborhood of
particles. We also allow the user to select multiple neighborhoods of particles for comparison.

\begin{figure}[t]
	\begin{center}
	\begin{subfigure}[b]{0.5\textwidth}
		\includegraphics[width=\linewidth]{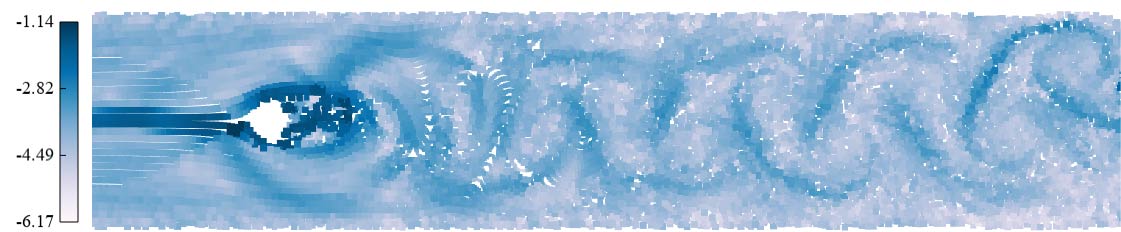}
		\caption{Diffusion Separation $s=7$}
	\end{subfigure}
	\begin{subfigure}[b]{0.5\textwidth}
		\includegraphics[width=\linewidth]{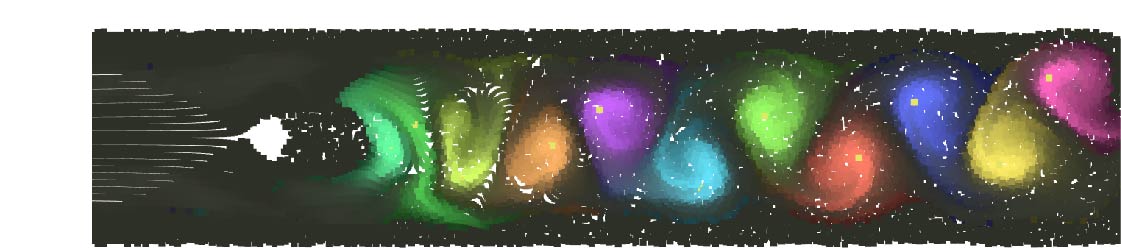}
		\caption{Similarities $s = 7$}
	\end{subfigure}
	\begin{subfigure}[b]{0.5\textwidth}
		\includegraphics[width=\linewidth]{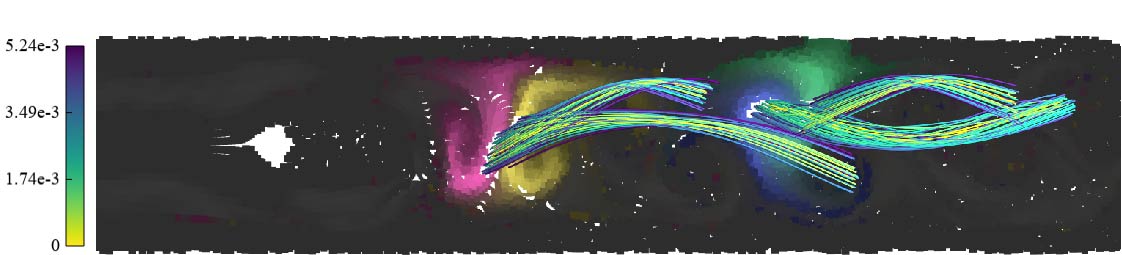}
		\caption{Particle Neighborhoods}
	\end{subfigure}
	\end{center}
	\caption{We show separation (a), similarity from a set of source particles (b), and their particle trajectory neighborhoods (c) for \textsf{2D Cylinder}.}
	\label{fig:2dcylinder}
\end{figure}

% example: 2D cylinder
\textbf{2D Similarity Example} We demonstrate our tools for exploring similarity by considering particles produced from a 2D fluid simulation, which we
denote \textsf{2D Cylinder}. We follow the experimental setup of~\cite{kanaris2011three} by placing a cylinder in the center of a 2D domain,
with boundary conditions prescribed by a Poiseuille, parabolic velocity profile, and a Reynolds number of 300. The fluid simulation software
Gerris~\cite{popinet2003gerris} is used to perform the simulation.
The result of the simulation is a von K\'{a}rm\'{a}n vortex street that
forms in the wake of the cylinder, and we would like to see if our similarity measure allows us to group the resulting vortices. We analyze particle
flows starting at time $t=10.6$ for a duration of $\tau = 1.8$ with $T=75$ time steps, and use a small scale of $s = 7$ in order
to capture the fine-grained vortex features. Figure~\ref{fig:2dcylinder}(a) shows our separation measure, highlighting the general shape of
the vortex street. Figure~\ref{fig:2dcylinder}(b) shows a set of distance fields that are formed by selecting particles enclosed by these separating
regions. We can observe the spatial extent of individual vortices through distance selection. Figure~\ref{fig:2dcylinder}(c) shows particle neighborhoods for each
of the selected particles. The neighborhoods illustrate the formation of particle groups in each vortex.

%Note that particles contain a strong translational component due to the initial velocity conditions of the simulation. However, since
%our method is objective, we can factor out this translation and extract the vortices using our distances.

% select particles based on separation - provide overview, then get details
\textbf{3D Particle Selection} In order to provide the user a useful means of selecting particles for similarity visualization in 3D, we guide selection by separation. We provide
two mechanisms for selection. The first is done by inverting the opacity mapping of separation through $(1-\hat{\gamma_s}^{a})$, which displays \emph{grouped} particles,
or particles that are dual to separation. These represent tight groups of particles that contain similar motion.
Through opacity mapping, the user can select grouped particles that are sufficiently opaque, as determined
by a predefined threshold. A limitation of this technique is that grouped particles tend to form large volumetric regions, thus occlusion can lead
to difficulties in performing selection.
On the other hand, separating particles form 1D and 2D structures that provide a summary of the data with reduced visual clutter, and tend to bound regions of grouped particles.
Hence we also use separating particles as a mechanism to select grouped particles.
For a selected separating particle that is sufficiently opaque, as determined by $\hat{\gamma_s}^{a}$, we select a nearby grouped particle.
Namely, we find the closest grouped particle that is in front of the separating particle, based on the camera viewpoint.
This allows the user to obtain an overview of the flow via separation, followed by a more detailed view of distances through particle selection.

% example: 3D steady ABC
We show 3D selection for the \textsf{ABC} flow. In Figure~\ref{fig:steadyabc}(b) the user selects grouped particles, and we show
their corresponding neighborhoods. Note that this depicts both individual groups, and how motion behaves within groups, i.e. swirling
motion in the ABC flow. Figure~\ref{fig:steadyabc}(c) shows color and opacity-mapped diffusion distances for these selected particles,
and one can observe that we recover the 6 main clusters of the \textsf{ABC} flow, as established from prior work~\cite{banisch2017understanding}.

\subsection{Particle Similarity and Coherent Sets}

For large scales, diffusion distances begin to resemble a coarse clustering of particles, and it can be shown~\cite{nadler2007fundamental}
that diffusion distances are related to spectral clustering from a random walks perspective.
As scale increases, the diffusion distances are mostly affected by the eigenvectors $\mb{u}^i$
with the largest eigenvalues, and it is these eigenvectors that are employed for spectral clustering. Spectral clustering of particles is a common technique
for computing coherent sets in flows~\cite{Hadjighasem16,banisch2017understanding}. For instance, in Hadjighasem et al. vortices
are extracted as coherent sets using spectral clustering, and background flow that does not comprise vortices is separated and considered incoherent~\cite{Hadjighasem16}.
Our technique can thus be used as a way to access coherent sets, but in an interactive manner that is driven by scale.
Interactively editing the scale allows the user to explore the strength of clusters, and thus provides for a soft notion of coherent sets modulated by scale.

\begin{figure}[!t]
	\begin{center}
	\begin{subfigure}[!t]{0.155\textwidth}
		\includegraphics[width=\linewidth]{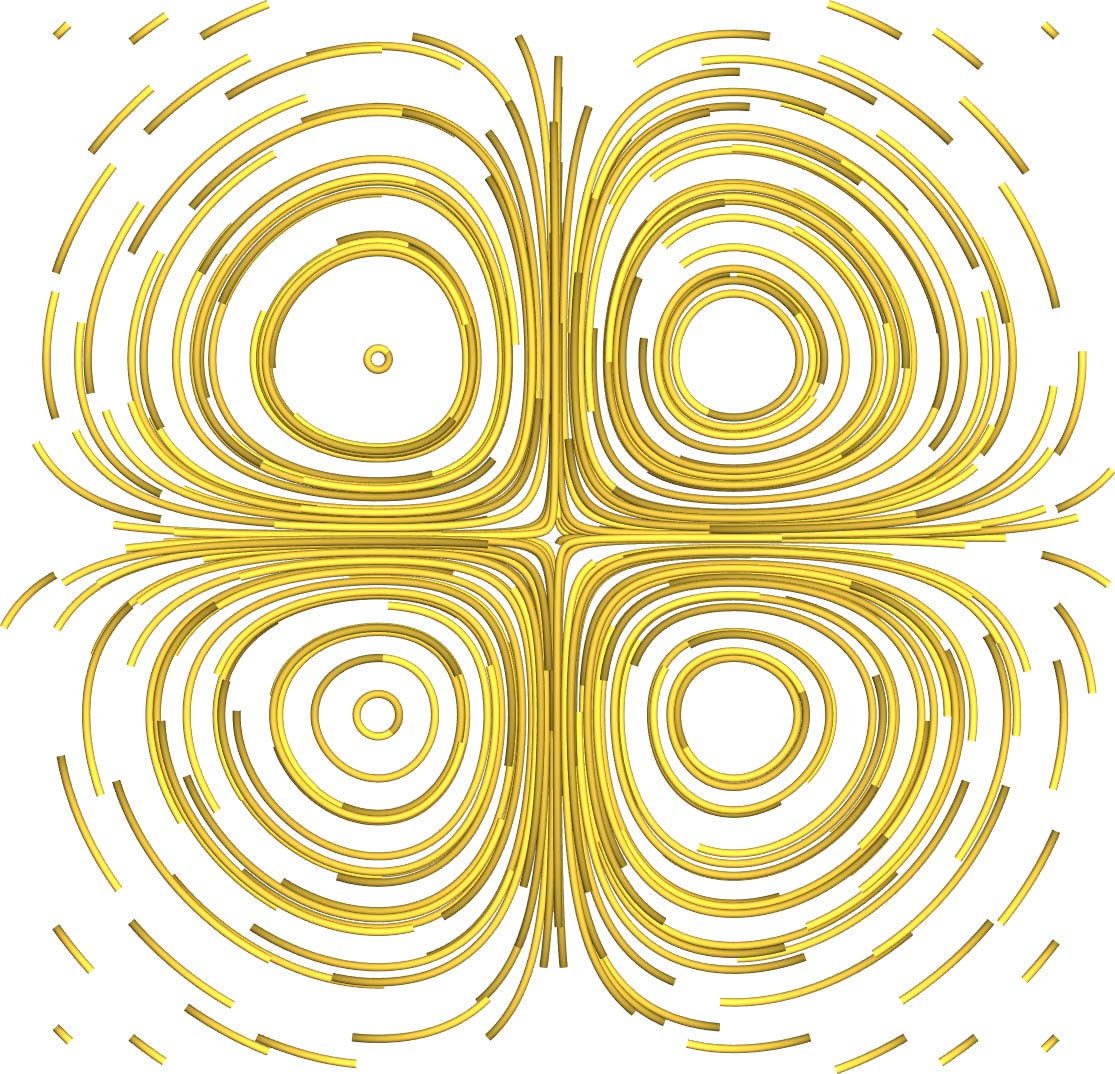}
		\caption{Particles}
	\end{subfigure}
	\begin{subfigure}[!t]{0.15\textwidth}
		\includegraphics[width=\linewidth]{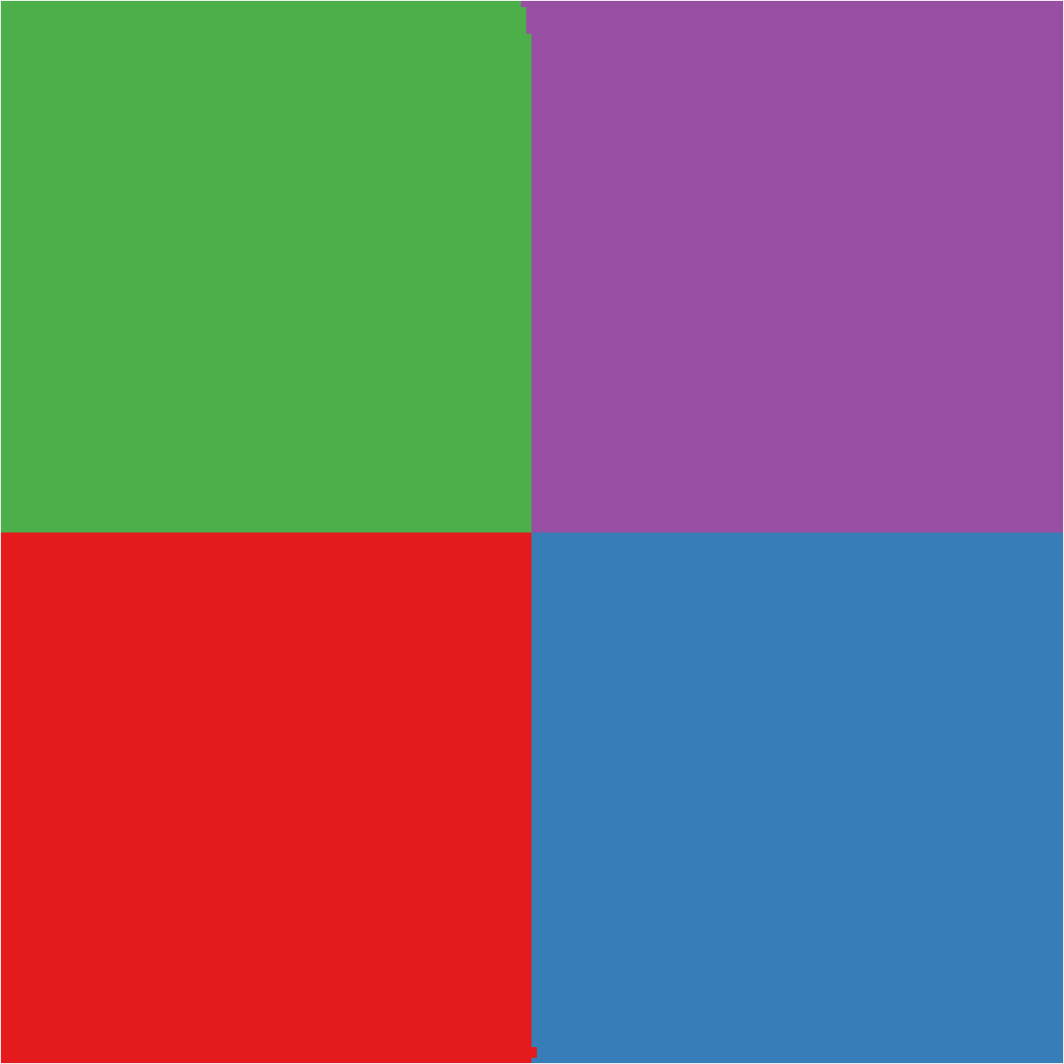}
		\caption{Clustering}
	\end{subfigure}
	\begin{subfigure}[!t]{0.15\textwidth}
		\includegraphics[width=\linewidth]{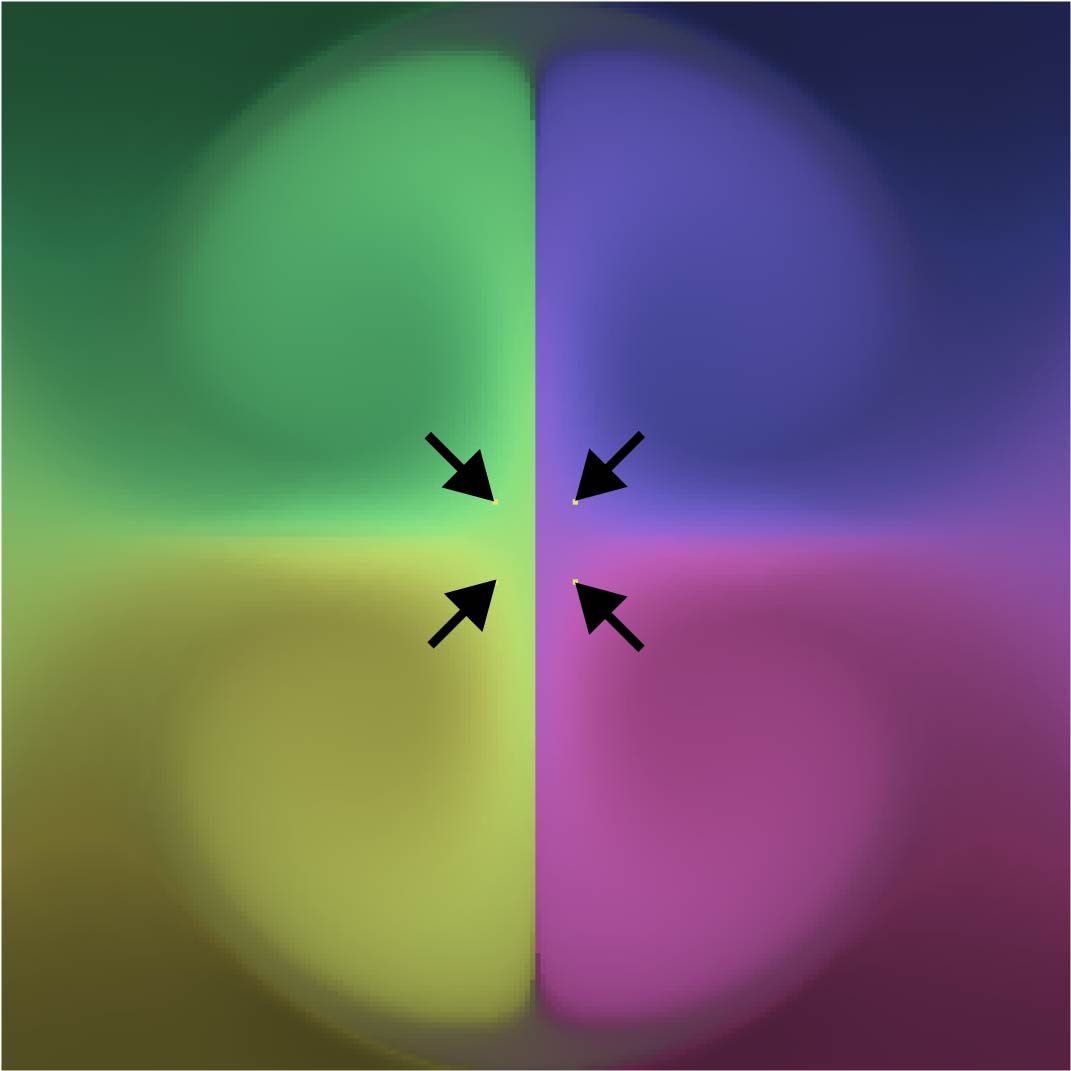}
		\caption{Distances}
	\end{subfigure}
	\end{center}
	\caption{We show particles in the \textsf{Four Centers} flow (a), the results of spectral clustering of particles (b), and diffusion distances (c) from
	four source particles indicated by arrows. Note how the distances induce a partition of the domain consistent with spectral clustering.}
	\label{fig:fourcenters}
\end{figure}

We illustrate this relationship with a steady vector field example, where a clustering of the domain corresponds to vector field topology. We use the \textsf{Four Centers} flow, comprised of four vortices which partition the domain,
and whose vector field is defined by:
\begin{equation}
	v(x,y)= \begin{pmatrix} -x \cdot e^{-x^2-y^2} \cdot (2 y^2 - 1)\\
	-y \cdot e^{-x^2-y^2} \cdot (2 x^2 - 1) \end{pmatrix}
\end{equation}
We uniformly seed particles in the domain $[-2,2]^2$, and compute particles by integrating the vector field starting at $t_1 = 0$
for a duration of $\tau = 10$, using $T = 500$ time steps. Figure~\ref{fig:fourcenters}(a) shows example particles, and results for spectral
clustering are shown in Figure~\ref{fig:fourcenters}(b), highlighting the four regions of uniform flow. Particles are visualized using diffusion distances from four
different source particles in Figure~\ref{fig:fourcenters}(c) for a scale of $s = 300$. We show particle positions at time $t_1$
and color map each particle based on its smallest distance to all source particles. We observe that particles grouped by diffusion distance coincide with cluster regions. Particle distance selection can thus
be used as a form of interactive clustering, where the user can select a particle and inspect the group of particles that have low distance to this particle.

\subsection{Objectivity of Similarity}

Our particle-based similarity is \emph{objective}, an essential property for meaningful user interaction.
Objectivity refers to invariance under smooth, time-varying rigid transformation coordinate changes~\cite{haller2005objective}
\begin{equation}
	\tilde{\mb{p}_t} = Q_t \mb{p}^i_t + \mb{b}_t,
\end{equation}
where $Q_t$ and $\mb{b}_t$ is the rotation and translation at time step $t$, respectively, and $\tilde{\mb{p}_t}$ represents $\mb{p}_t$
undergoing the transformation. The particle diffusion geometry is objective due to the distance $d_E$ used in the kernel function $k$. The distance $d_E$ is strictly
a function of Euclidean distance between particle positions, which is invariant to rigid transformations. Since the remaining steps of
constructing diffusion geometry only make use of the kernel matrix $K$, it follows that the diffusion geometry is objective.

\begin{figure}[!t]
	\begin{center}
	\begin{subfigure}[!t]{0.15\textwidth}
		\centering
		\includegraphics[width=\linewidth]{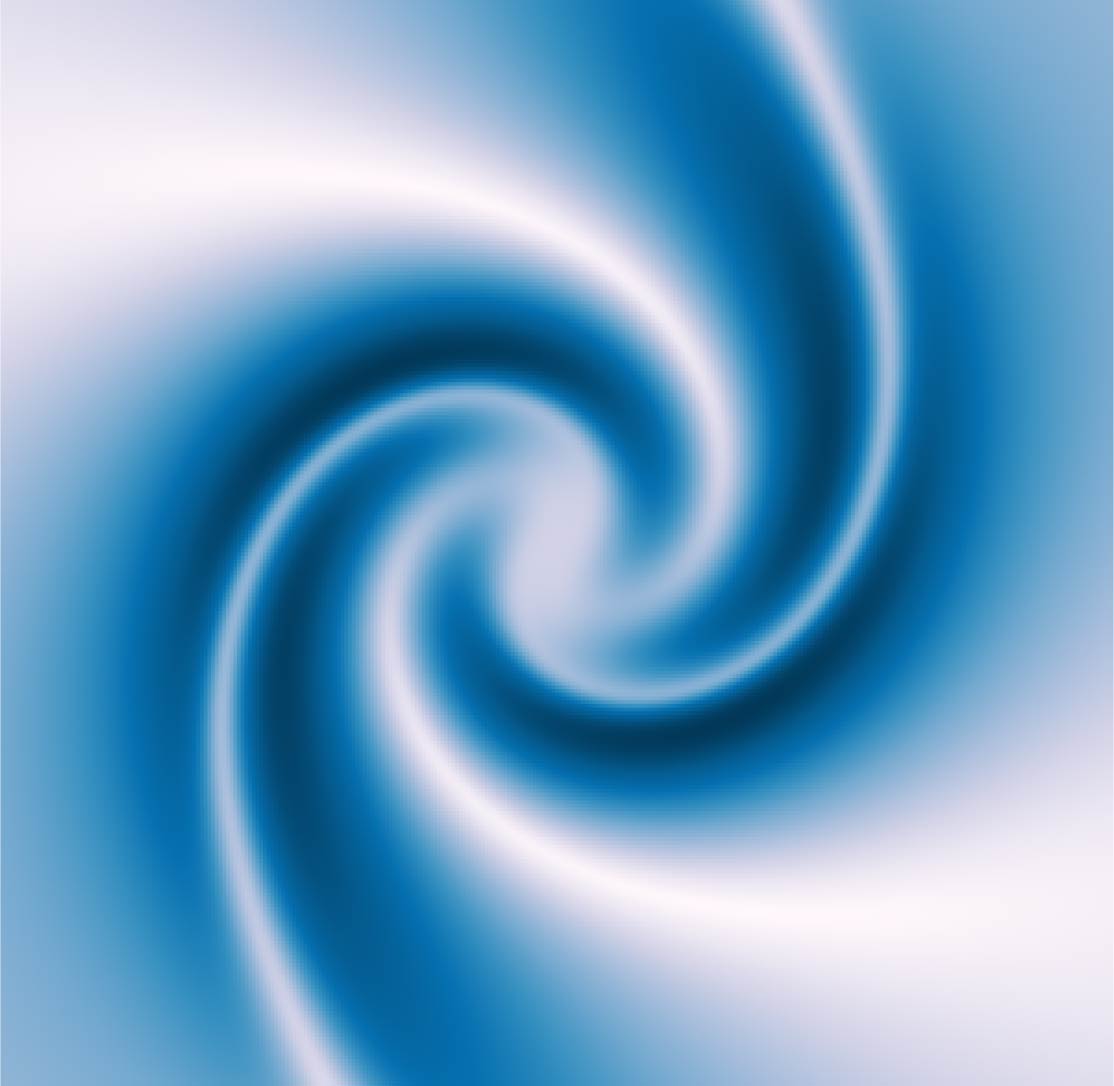}
		\caption{FTLE}
	\end{subfigure}
	\begin{subfigure}[!t]{0.15\textwidth}
		\centering
		\includegraphics[width=\linewidth]{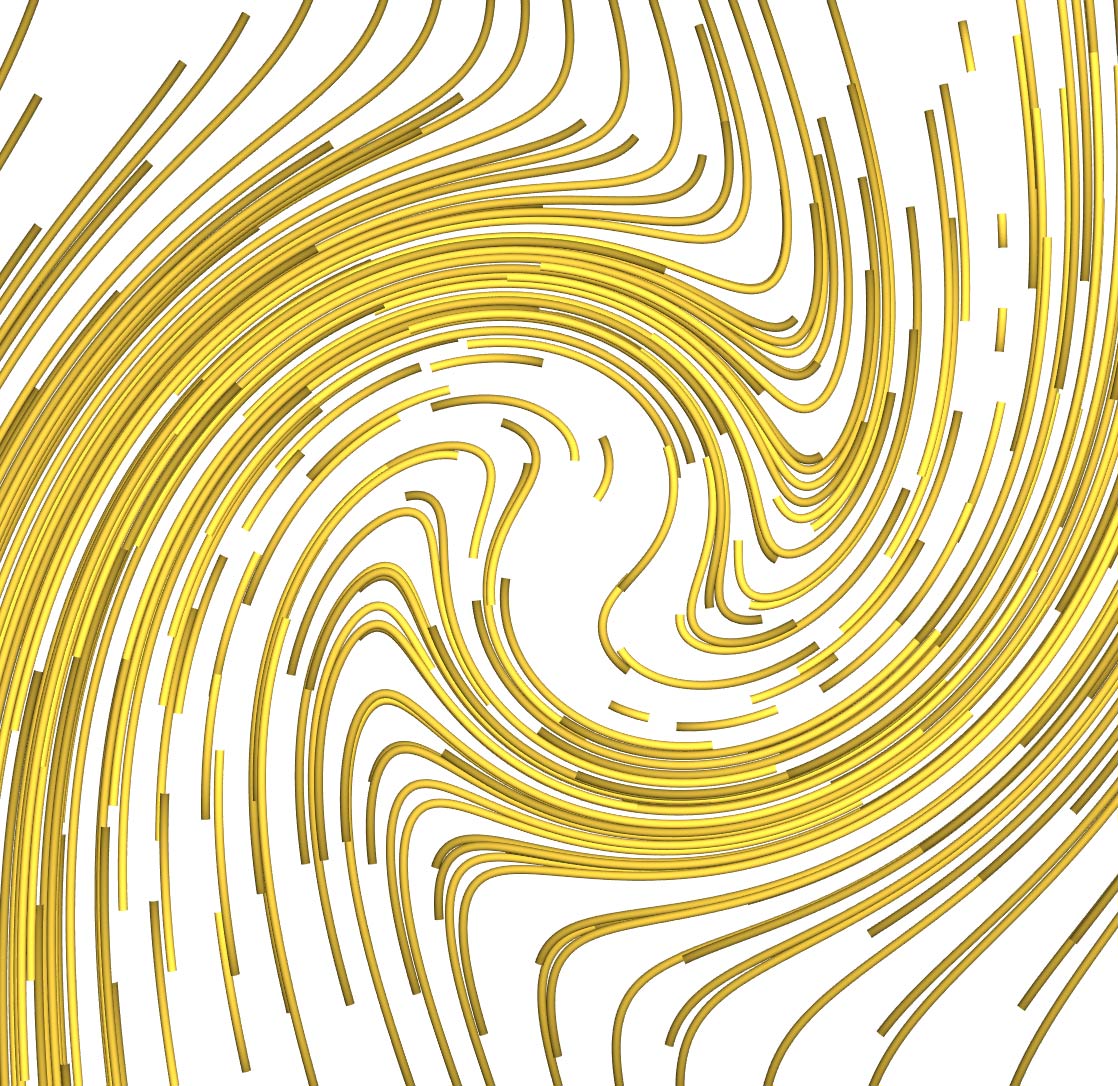}
		\includegraphics[width=\linewidth]{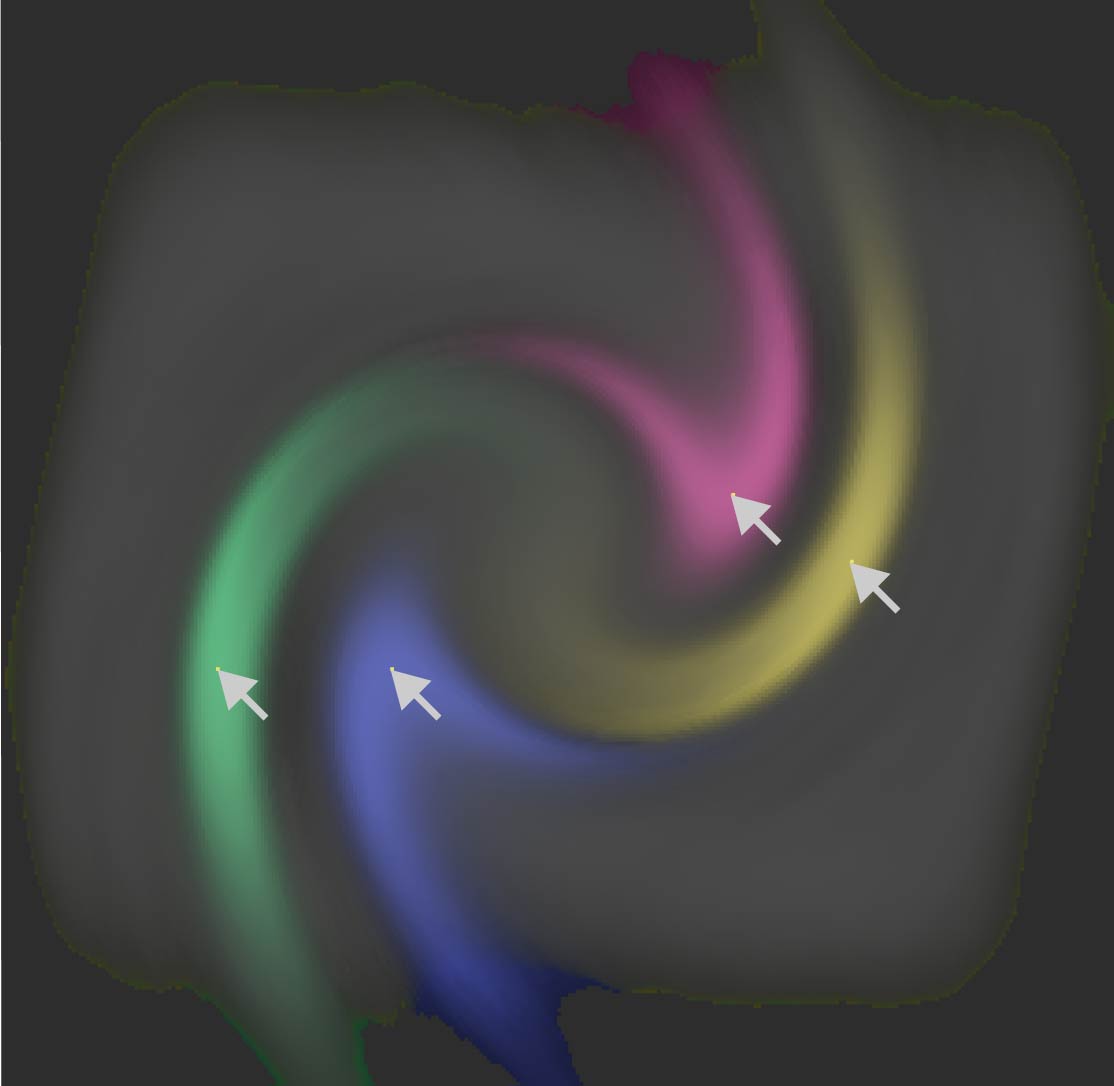}
		\caption{Original Particles}
	\end{subfigure}
	\begin{subfigure}[!t]{0.15\textwidth}
		\centering
		\includegraphics[width=\linewidth]{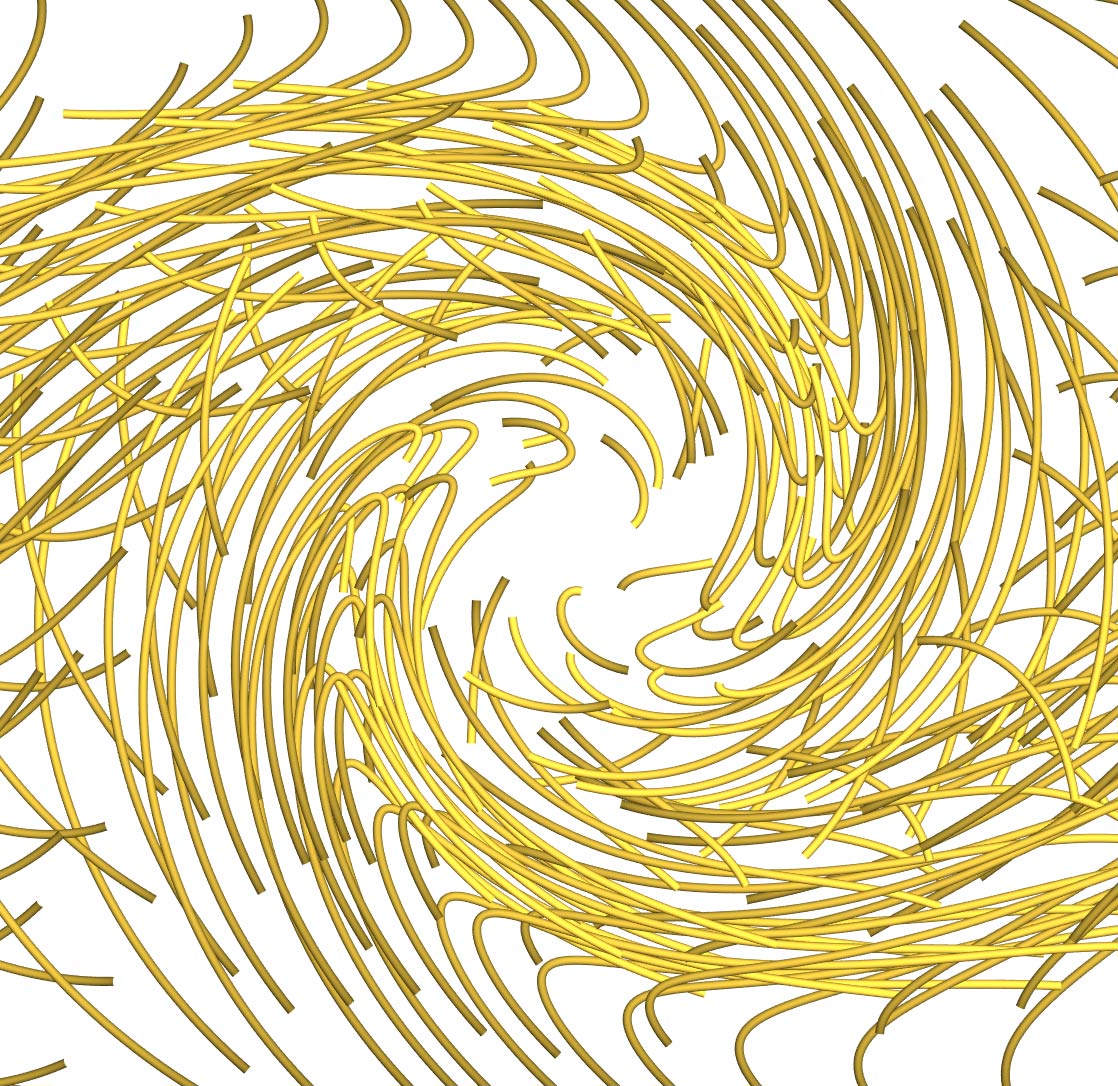}
		\includegraphics[width=\linewidth]{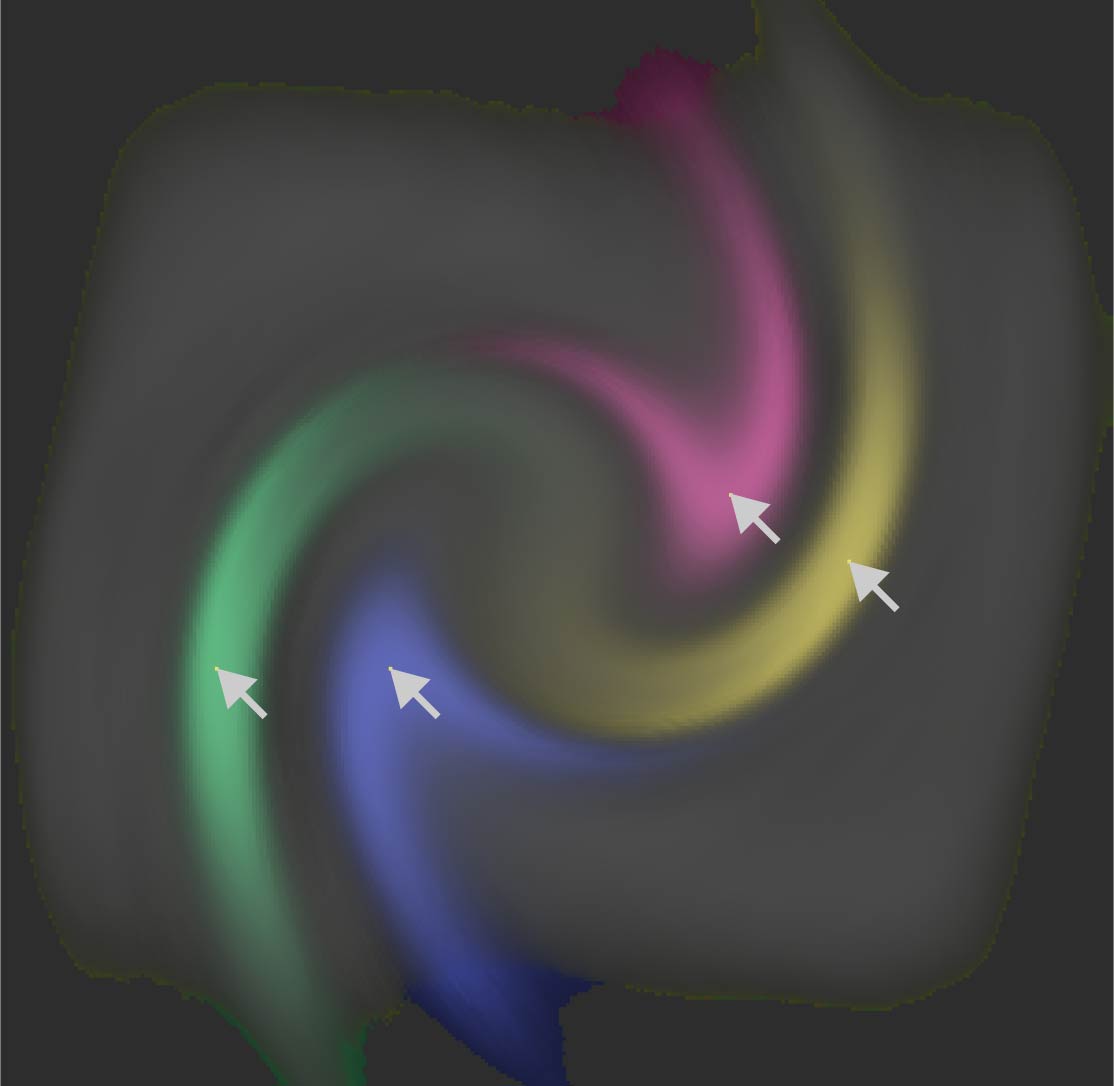}
		\caption{Transformed}
	\end{subfigure}
	\end{center}
	\caption{For the \textsf{Spiral Focus} flow we show its t-FTLE  (a), particles and corresponding distances (b), as well as
	the particles undergoing a time-varying rotation, and the resulting distances (c). Note the distances remain unchanged
	when the particles undergo such a coordinate change.}
	\label{fig:objectivity}
\end{figure}

Objectivity is a useful property to satisfy when a coordinate change of the flow can obscure the flow's features, in particular vortices~\cite{gunther2017generic}. We illustrate
this through an example using an analytical flow map, namely the \textsf{Spiral Focus} flow of Kuhn et al.~\cite{kuhn2012benchmark}.
This dataset is characterized by swirling motion that causes flow divergence, with one set of flow moving towards the center,
and another set moving away. We consider particles uniformly seeded in the domain $[-4,4]^2$ starting at time $t_1 = 0$ for a duration of $\tau = 2.5$,
using $T = 100$ time steps, see Figure~\ref{fig:objectivity}(b)-top for example particles. Figure~\ref{fig:objectivity}(b)-bottom shows color-mapped distances for particles at $t_1$, where we
select four different source particles from which to compute distances. Note how the distances respect the separation ridges found in
the t-FTLE  in Figure~\ref{fig:objectivity}(a). We perform a rotation of the flow map by prescribing a rotating coordinate change with angular velocity of $25$.
Figure~\ref{fig:objectivity}(c)-top shows example particles from this transformation and Figure~\ref{fig:objectivity}(e)-bottom shows distances from the same source particles. We find these diffusion
distances are the same as the distances of the original particles.

The \textsf{2D Cylinder} example of Figure~\ref{fig:2dcylinder} further emphasizes the benefit of objectivity for visualizing particle similarity.
In this example, particles contain a strong translational component due to the initial velocity conditions of the simulation. Due to objectivity,
similarity neighborhoods correspond to a reference frame where particles are not advected by this translation, and thus the selected clusters correspond
to the vortices shed behind the cylinder.

\section{Results}

We have applied our technique to a variety of datasets, see Table~\ref{table} for dataset statistics, parameters used for each dataset, and computational timings in Table~\ref{table}.

\begin{table*}[!t]
\begin{center}
	\begin{tabular}{| c | c | c | c | c | c | c | c |}
		\hline
		name & num particles & num landmarks & time steps & $\alpha$ & landmarks (s) & kernel (s) & eigendecomposition (s) \\ \hline
		\textsf{Double Gyre} & 7.2K & 1K & 100 & 1 & 0.19 & 0.22 & 5.4 \\ \hline
		\textsf{2D Cylinder} & 19.4K & 5K & 75 & 1 & 2.92 & 0.98 & 21.3 \\ \hline
		\textsf{Sine Ridge} & 40K & 5K & 100 & 0.75 & 5.53 & 1.62 & 35.7 \\ \hline
		\textsf{Four Centers} & 40K & 5K & 500 & 0.75 & 5.96 & 0.93 & 36.1 \\ \hline
		\textsf{Spiral Focus} & 90K & 5K & 100 & 0.75 & 12.9 & 1.91 & 37.83 \\ \hline
		\textsf{ABC} & 262K & 5K & 40 & 1.25 & 56.3 & 30.4 & 52.5 \\ \hline
		\textsf{Heated Cylinder} & 264K & 15K & 75 & 0.85 & 104.8 & 10.1 & 110.0 \\ \hline
		\textsf{Dark Sky} & 500K & 25K & 98 & 0.9 & 497.0 & 99.2 & 208.1 \\ \hline
		\textsf{Cloud Collapse} & 837K & 25K & 54 & 0.9 & 544.5 & 77.5 & 192.6 \\ \hline
		\textsf{3D Cylinder} & 1M & 25K & 25 & 0.9 & 707.7 & 118.9 & 243.6 \\ \hline
		\textsf{Dust Settling} & 1M & 25K & 25 & 1.0 & 627.7 & 119.1 & 282.2 \\ \hline
	\end{tabular}
\end{center}
	\caption{Dataset statistics, parameters used, and computational timings for experiments. The $\alpha$ parameter is the bandwidth scale used in Eq.~\ref{eq:bandwidth}.
	The last three columns are timings, in seconds, for computing landmarks, computing the kernel matrix $K$, and performing the eigendecomposition, respectively.}
	\label{table}
\end{table*}

\begin{figure}[!t]
	\centering
	\begin{tikzpicture}[thick,scale=0.925, every node/.style={transform shape}]
	\node[inner sep=0pt,anchor=south west]  at (0,0)
	{\includegraphics[width=1.0\linewidth]{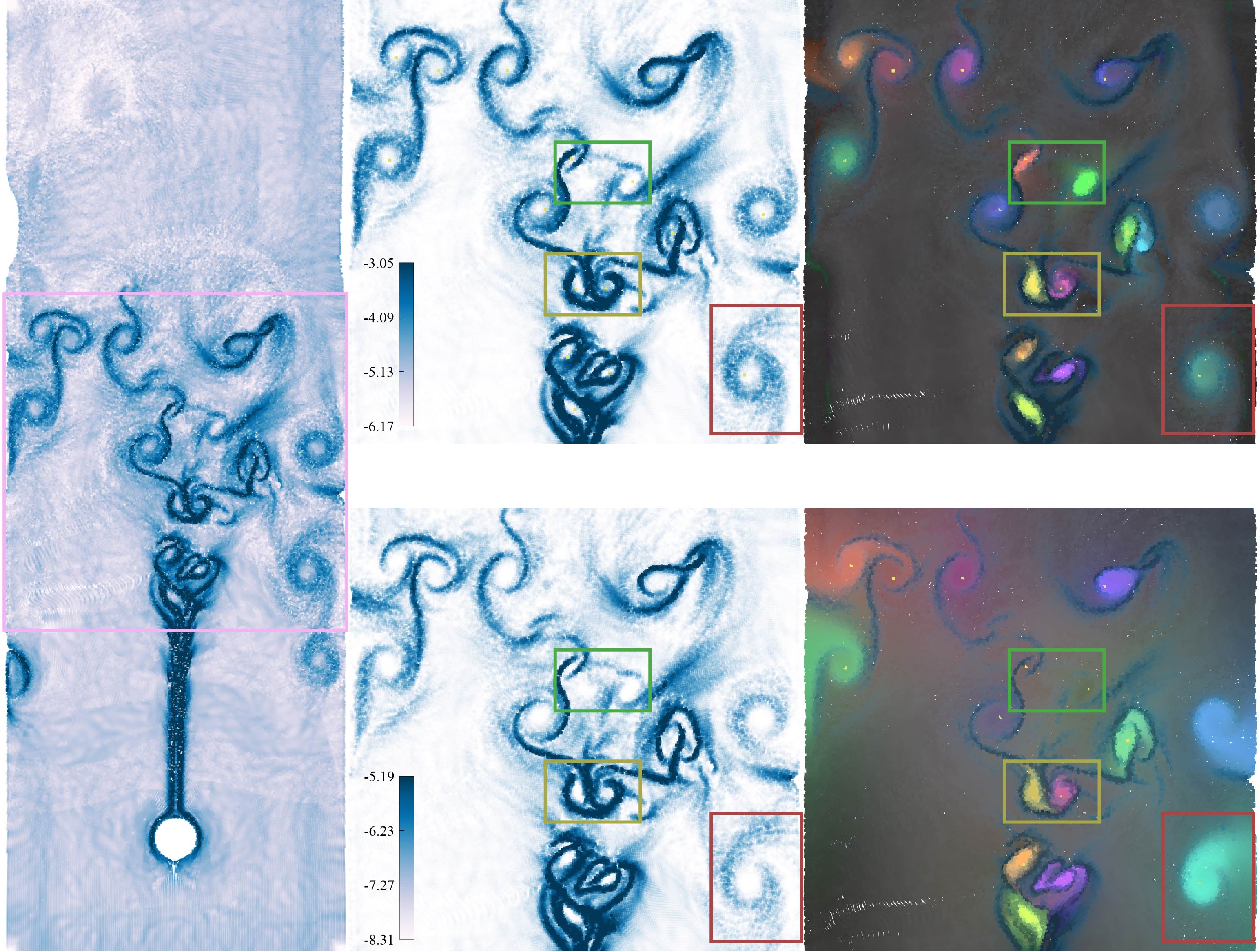}};
	\node at (5.5,3.4) {(a) Diffusion Similarities $s = 10$};
	\node at (5.5,-0.25) {(b) Diffusion Similarities $s = 50$};
	\end{tikzpicture}
	\caption{We show multi-scale analysis of the \textsf{Heated Cylinder} flow. We show our diffusion separation field in addition to a set of user-selected
	source points and their diffusion distance neighborhoods for two different scales.}
	\label{fig:boussinesq}
\end{figure}

\textbf{Implementation Details} As discussed in Section~\ref{sec:markov}, the kernel matrix $K$ (Eq.~\ref{sec:markov}) is sparse, enabling us to assemble $K$ by performing
neighborhood queries. However, performing high-dimensional neighorhood queries using time-dependent bandwidths (Eq.~\ref{eq:bandwidth}) is expensive when the number of particles is large.
For efficient computation of neighborhoods we
use the positions at the starting time step $t_1$ to search for a particle's candidate neighbors, and then compute $d_E$ over all
time steps for these neighbors. This approximation avoids the costly operation of performing neighbor queries in the high-dimensional space as determined by $d_E$, and results in $K$ being a subset
of the neighbors computed using $d_E$.
However, in practice we find that through diffusion particles that are not initially neighbors due to this approximation become similar for small diffusion scales.
Last, the computation of covariance-based separation in Eqs.~\ref{eq:particlecov} and~\ref{eq:diffusioncov} requires the definition of a spatial neighborhood at each particle.
For simplicity, we use the $k$-nearest neighbors to each particle where $k=9,27$ for $d=2,3$, respectively.

\subsection{Experimental Results}

\textbf{Heated Cylinder} In this example we consider 2D flow simulation of convection from a heated cylinder.
This results in a turbulent plume
and the formation of vortices nonuniformly distributed throughout the domain. We consider particle flows that start at $t_1 = 16.6$ with an integration duration
of $\tau = 1.3$, and uniformly sample $T = 75$ time steps. Gerris~\cite{popinet2003gerris} is used to perform the simulation.

\begin{figure}[!t]
	\begin{tikzpicture}
		\node[inner sep=0pt,anchor=south west]  at (0,0)
		{\includegraphics[width=1.0\linewidth]{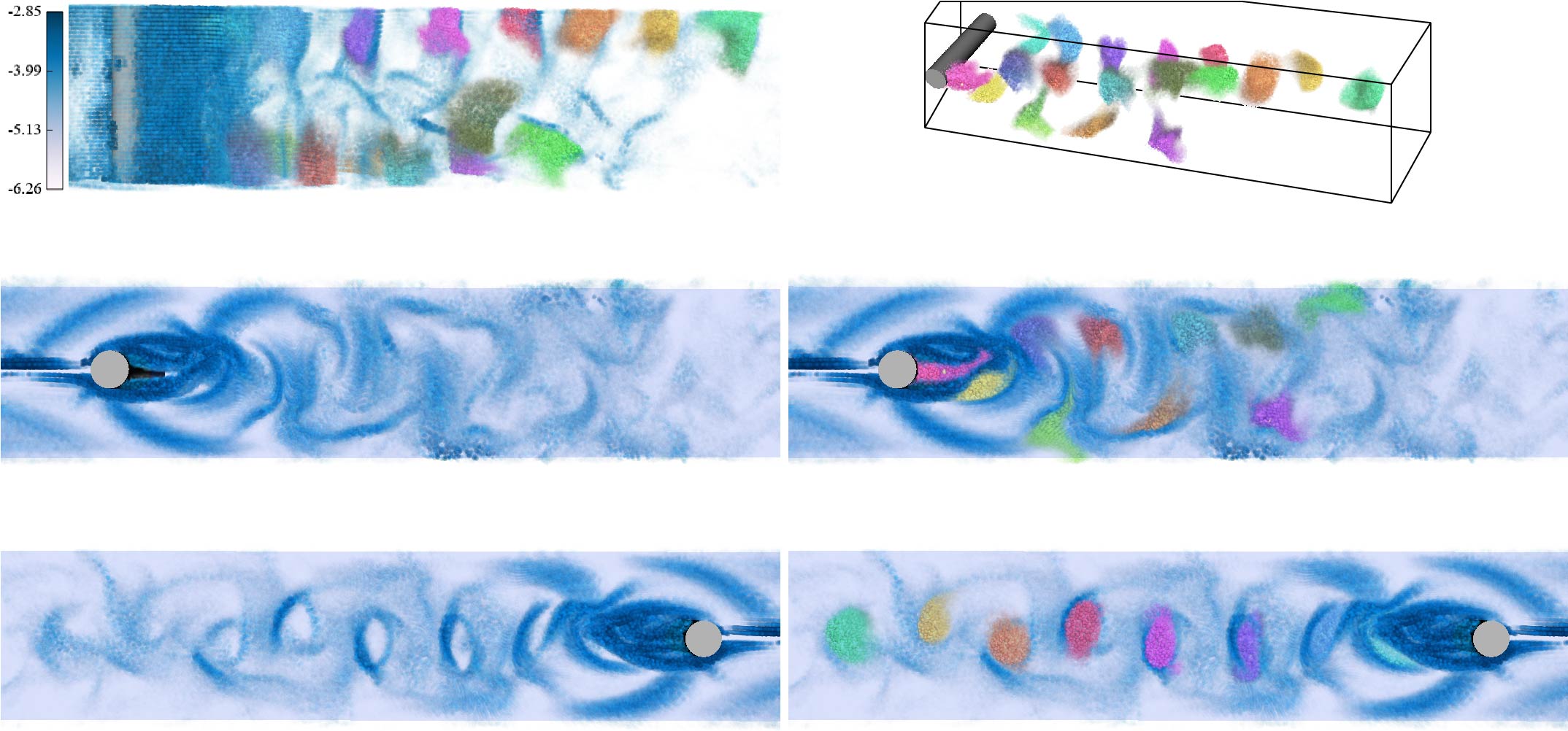}};
		\node at (2.35,2.83) {(a) Top-Down View};
		\node at (6.7,2.83) {(b) Unoccluded Neighborhoods};
		\node at (2.3,1.3) {(c) Separation Side View 1};
		\node at (6.7,1.3) {(d) Similarity Side View 1};
		\node at (2.3,-0.2) {(e) Separation Side View 2};
		\node at (6.7,-0.2) {(f) Similarity Side View 2};
	\end{tikzpicture}
	\caption{We show results on \textsf{3D Cylinder} at different views of the domain for $s=14$, visualizing separation structures in (c) and (e), similarity
	in (b), and their combined visualizations in (a), (d), and (f). Note that the shapes of the neighborhoods suggest the non-2D flow behavior of the shed vortices.}
	\label{fig:cylinder3d}
\end{figure}
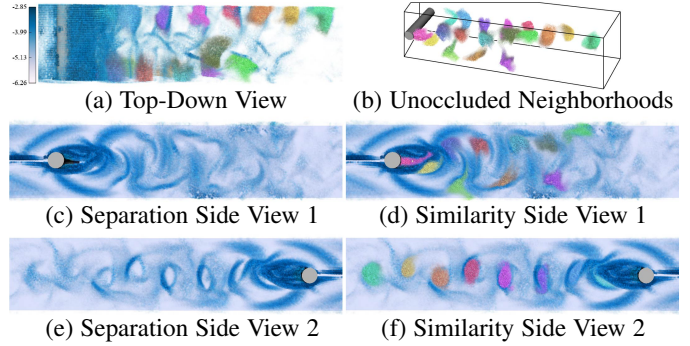

\begin{figure}[t]
	\begin{center}
	\includegraphics[width=0.9\linewidth]{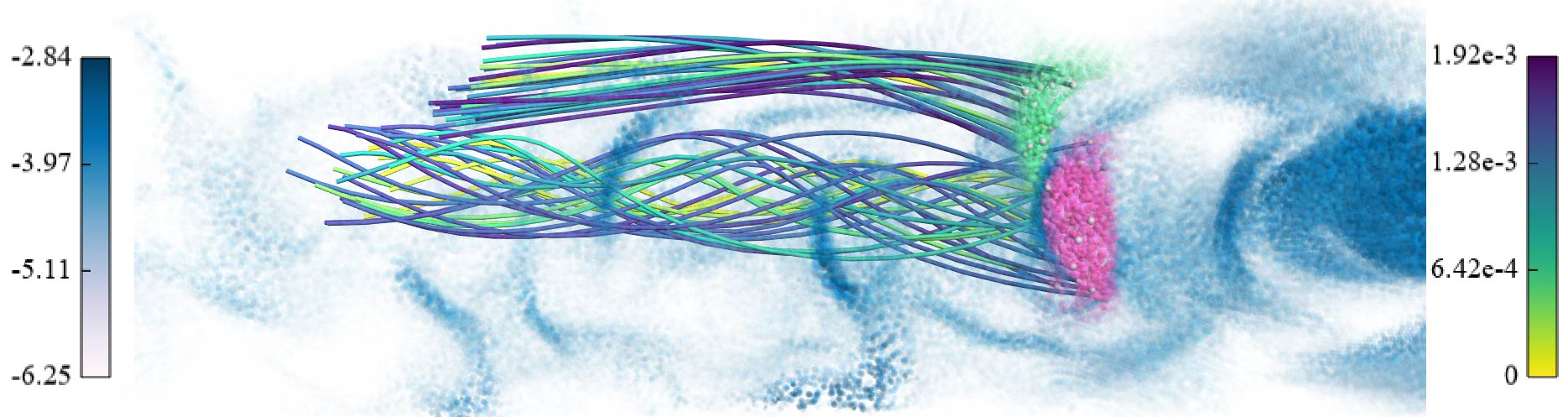}
	\end{center}
	\caption{For the \textsf{3D Cylinder} dataset we select neighborhoods on opposite sides of a ridge, and show particles in each of these neighborhoods. Note that swirling
	flow is demonstrated for particles in the shed vortex of the red neighborhood, while laminar flow is shown in the green neighborhood.}
	\label{fig:cyl3dtraj}
\end{figure}

Figure~\ref{fig:boussinesq} shows the results, where we highlight separation, and superimpose separation with user-selected similarity neighborhoods for two different scales.
Vortices can be seen from the separation boundaries, while the individual neighborhoods are typically bounded by the separation, and help to convey
the shape of the vortices. By analyzing the separation from small scale (a) to large scale (b) we can observe the strength of certain vortices. For instance,
the vortex highlighted in the red box breaks apart at its top boundary as we increase the scale. This is supported by its neighboring particles at the large scale, where
they expand out of this region. We can also observe how vortices that are close in Euclidean distance can blend together as scale increases, highlighted in green,
while other nearby vortices retain their neighborhood, highlighted in yellow. This suggests that the latter vortices are stronger than the former, in
that on the particle manifold it takes a shorter distance to walk between the vortices highlighted in green, compared to those highlighted in yellow.

\textbf{Flow Over 3D Cylinder} This simulation is a 3D analogue to the \textsf{2D Cylinder} simulation discussed in Figure~\ref{fig:2dcylinder}, following
the experimental setup of Kanaris et al.~\cite{kanaris2011three}. A 3D cylinder is placed in the center of the domain, fully occupying the spanwise direction.
The Reynolds number is set to 300. We consider particle flows starting at $t_1 = 9.6$ with an integration duration of $\tau = 2.3$ for $T = 25$ time steps,
and use Gerris~\cite{popinet2003gerris} to run the simulation.

Figure~\ref{fig:cylinder3d} shows the separation field and neighborhoods of user-selected particles at scale $s=14$, with views showing both sides of the domain,
as well as the top of the domain. For visual clarity, in (c)-(d) and (e)-(f) we filter out particles for which $z > 0$ and $z < 0$, respectively.
As was studied in~\cite{kanaris2011three}, we similarly obtain vortex shedding that is nonuniform in the spanwise direction, suggesting
that the flow is not merely two-dimensional. For instance note that on one side (e)-(f) the vortices are thinner compared to (c)-(d). The top-down view (a) and the
unoccluded set of particle neighborhoods (b) shows the distribution of the vortices over the domain, confirming this variation.

Figure~\ref{fig:cyl3dtraj} provides a closer inspection on separation. We select two particles on opposite sides of a separation sheet, where a vortex exists
on one side but does not exist on the other side. We show particles seeded in these two neighborhoods, showing how separation occurs: within the enclosed vortex identified
by the red neighborhood, particles contain swirling flow whereas those not within a vortex, the green neighborhood, are of laminar flow. 
Viewing the particle paths, as in the \textsf{2D Cylinder}, shows how our method is objective, in that despite the strong translational component to the particle flow, we can
still group particles based on their swirling motion.

\textbf{Dust Settling} We analyze data from astrophysical fluid dynamical systems and the formation of protoplanets. We consider the problem
studied in Lor{\'e}n-Aguilar \& Bate~\cite{loren2015toroidal} which investigates the impact of a baroclinic instability, or the misalignment of gradient pressure from density pressure,
on dust settling. In this SPH simulation, particles are initially distributed inside of a torus and rotate around the torus' gravitational center,
while the mixing of gas and dust particles results in the dust particles to slowly move towards the center. It was shown in~\cite{loren2015toroidal} that baroclinic instability
results in toroidal vortices. This process can potentially inhibit planet formations.

\begin{figure}[t]
	\begin{center}
	\begin{subfigure}[b]{0.5\textwidth}
		\includegraphics[width=\linewidth]{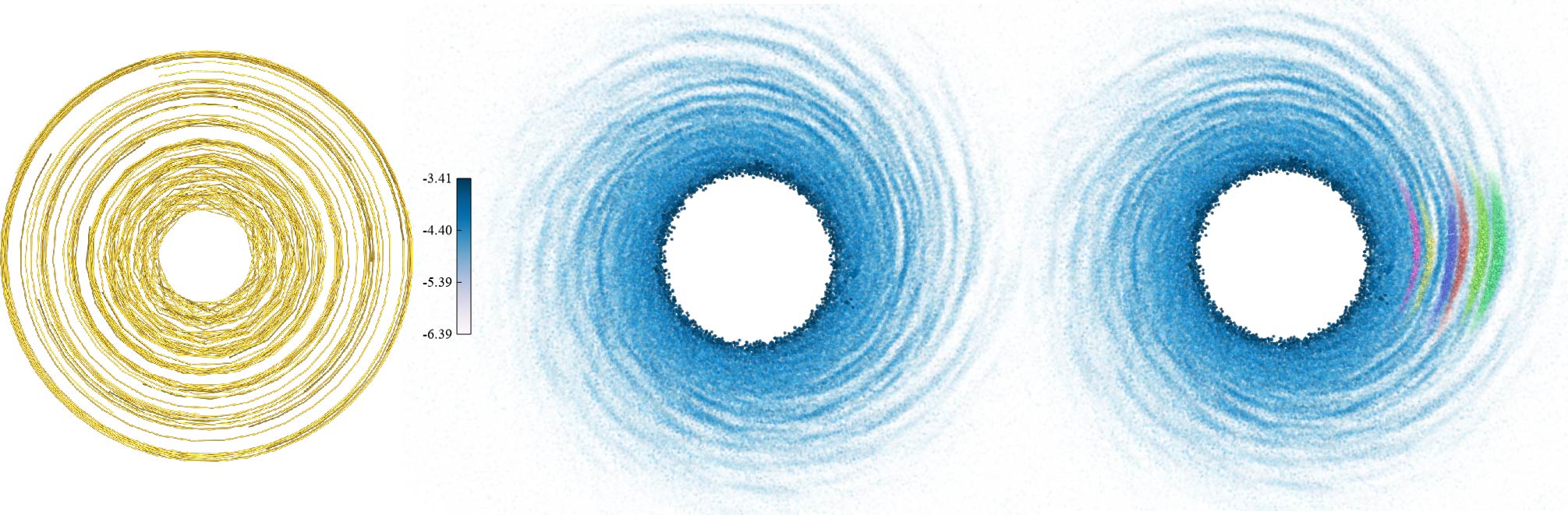}
		\caption{Separation and Similarity $s = 50$}
	\end{subfigure}
	\begin{subfigure}[b]{0.5\textwidth}
		\includegraphics[width=\linewidth]{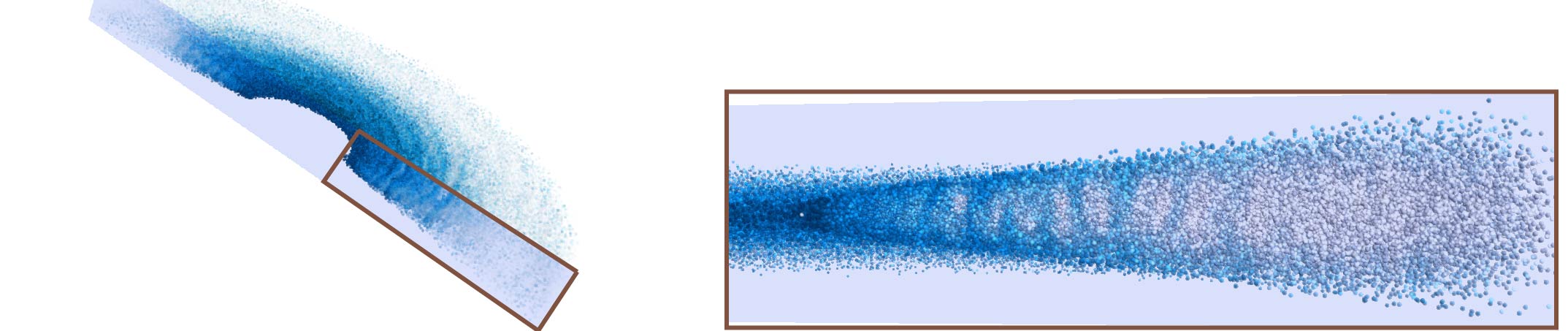}
		\caption{Separation Side View}
	\end{subfigure}
	\end{center}
	\caption{We show particle separation and similarity for dust settling in the formation of protoplanets (a), along with a cut-away side view (b). Despite the
	dominant rotational motion exhibited by the particles in (a)-left, we are still able to extract ring-like structures in the separation. The side view (b) shows
	how separation regions progress radially outward from the torus center.}
	\label{fig:toroidalvortices}
\end{figure}

We analyze dust particles to observe if their motion also suggests a baroclinic instability. We have taken particles produced via the simulation starting from $t_1 = 631$
for duration $\tau = 47$ and sampling $T = 25$ time steps.
Figure~\ref{fig:toroidalvortices}(a) (left) shows example particles over this temporal range as viewed from above. Particles rotate around the center at different
angular velocities, where particles closer to the center move at a higher velocity. Over time this results in the outward propagation of particles from the center and
the formation of ring structures. Our diffusion separation measure captures individual rings as shown in Figure~\ref{fig:toroidalvortices}(a) (right),
while the similarity neighborhoods show how particles are grouped in their motion. We find that particle separation is correlated with dust density (c.f.~\cite{loren2015toroidal} Figure 3).
Inspecting a cross section of the torus in Figure~\ref{fig:toroidalvortices}(b), we can observe additional structure in the separation. Formation of separation occurs in a manner that is related
to the formation of the baroclinic instability, and the presence of toroidal vortices (c.f.~\cite{loren2015toroidal} Figures 1,2).

\textbf{Dark Sky} We consider flows produced from the Dark Sky~\cite{skillman2014dark} dataset, a cosmology simulation of the large-scale universe
in which particles expand and form into galaxies. The expansion results in particles forming dense and compact clusters, curve-like filaments, surface-like sheets, and empty voids.
Previous visualization techniques have considered how to efficiently render the large amount of particles in such simulations~\cite{schatz2016interactive} and summarize filament structures
via topological methods~\cite{shivashankar2016felix}. In contrast, we use our technique to visualize the evolution of particle expansion and the resulting formation of clusters.
We use the 2M particle dataset from the SciVis 2015 Contest~\cite{warren_2014_10777} for $t_1 = 0$, a simulation duration of $\tau = 98$, number of sampled time steps $T = 99$,
and use a 500K subsampled set of particles for analysis and visualization.

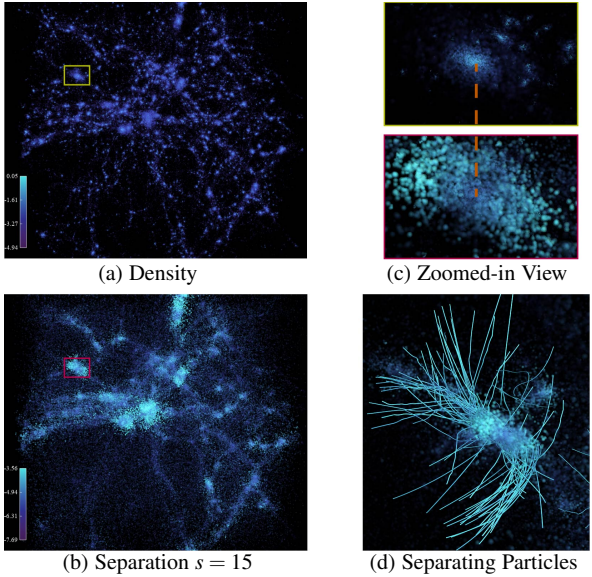
\begin{figure}[t]
	\centering
	\begin{tikzpicture}[thick,scale=0.875, every node/.style={transform shape}]
		\node[inner sep=0pt,anchor=south west]  at (0,0)
		{\includegraphics[width=1.0\linewidth]{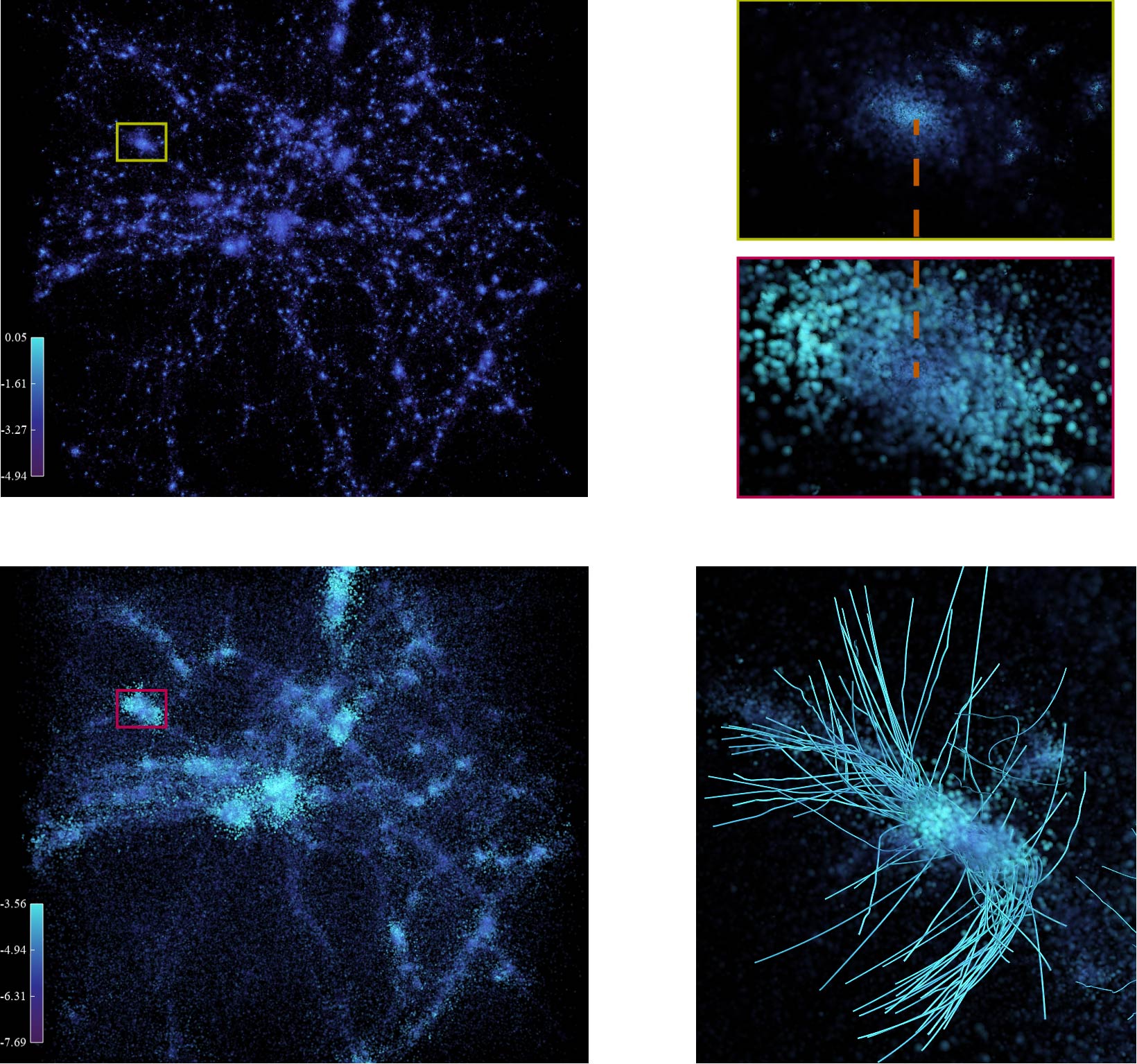}};
		\node at (2.2,4.2) {(a) Density};
		\node at (2.35,-0.2) {(b) Separation $s=15$};
		\node at (7.2,4.2) {(c) Zoomed-in View};
		\node at (7.15,-0.2) {(d) Separating Particles};
	\end{tikzpicture}
	\caption{For the \textsf{Dark Sky} dataset at the last time step we show how dense clusters (a) are surrounded by particles of attraction (b), or particles
	that have high backward-time separation. This is further emphasized in the zoomed region (c), and the set of highly attracting particles (d) demonstrate how the boundary was formed.}
	\label{fig:ds-sep}
\end{figure}

In Figure~\ref{fig:ds-sep} we illustrate how our method complements density-based visualization for particle positions at the last time step $t_T$.
Figure~\ref{fig:ds-sep}(a) shows the log-density field $d$ of the particles at $t_T$, computed by a $k$-nearest neighbor density estimate~\cite{biau2011weighted}:
\begin{equation}
	d(\mb{p}^i_{t_T}) = \log \left( \sum_{j \in N_i} \frac{1}{\lVert \mb{p}^i_{t_T} - \mb{p}^j_{t_T} \rVert} \right),
\end{equation}
where $N_i$ is the set of $\mb{p}^i$'s nearest neighbors at time $t_T$, and $k$ is set to 27.
Figure~\ref{fig:ds-sep}(b) shows our separation measure where we use backward-time separation in Equation~\ref{eq:diffusioncov}, such that diffusion covariance uses spatial neighborhoods at time $t_T$.
Note that a particle at $t_T$ has high separation measure if particles in its spatial neighborhood at $t_T$
have large variation in their diffusion geometry, indicative of diverging particle behavior. Spatially-close particles at $t_T$ of high separation thus
suggests particle attraction, where particles may have different origins, or more generally diverge in flow over all time steps.
We find that the presence of clusters (a) is correlated with attracting particles (b), in that particle neighborhoods gradually break up over time
in order to form tight, dense clusters, and separating particles form cluster boundaries. This is further highlighted in Figure~\ref{fig:ds-sep}(c),
where we show a zoomed-in cluster's density and separation, respectively. Furthermore, we show particle trajectories in Figure~\ref{fig:ds-sep}(d) for these
high-separation particles, where the per-time step bounds of particle positions are rescaled for visual clarity. For the particular zoomed-in cluster we can observe
that there are two main groups of particles that diverged to form its boundary.

\begin{figure}[t]
	\centering
	\begin{tikzpicture}[thick,scale=0.875, every node/.style={transform shape}]
		\node[inner sep=0pt,anchor=south west]  at (0,0)
		{\includegraphics[width=1.0\linewidth]{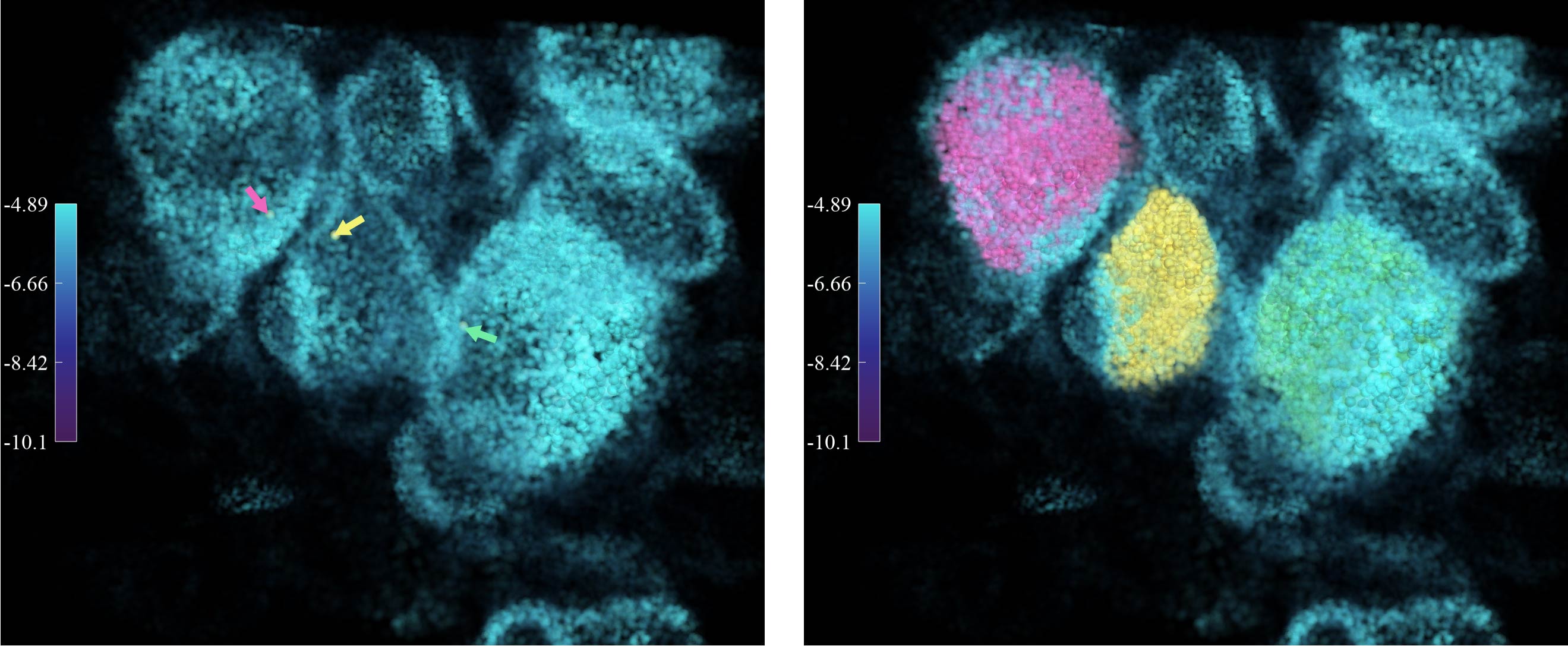}};
		\node at (2.25,-0.2) {(a) Separation $s=85$};
		\node at (6.8,-0.2) {(b) Similarity Neighborhoods};
	\end{tikzpicture}
	\caption{We show particle repulsion in \textsf{Dark Sky} through forward-time separation at the start of the simulation (a), highlighting particles for similarity neighborhood selection. The neighborhoods
	(b) are contained within the separating regions, suggesting their formation into dense clusters as the simulation progresses.}
	\label{fig:ds-sim}
\end{figure}

Figure~\ref{fig:ds-sim} depicts a different perspective of \textsf{Dark Sky} by showing particle separation at the first time step $t_1$, where separation indicates particle repulsion.
We observe that particle repulsion leads to the formation of sheet-like surfaces in Figure~\ref{fig:ds-sim}(a) that envelope
low separation voids. By inspecting similarity neighborhoods at user-specified positions in these voids in Figure~\ref{fig:ds-sim}(b), we can see that particle
neighborhoods expand to the separation boundaries. This behavior suggests how particles from the first time step form cluster centers and cluster boundaries. Namely,
for each similarity neighborhood its comprised set of particles flow in a similar and coherent way, and as we show in Figure~\ref{fig:ds-sep} cluster boundaries
manifest as high separation regions, thus these similarity neighborhoods are likely to form dense and compact clusters at $t_T$.

\subsection{Landmark Evaluation}
\label{sec:landeval}

As described in Section~\ref{sec:landmark}, we have used temporally-subsampled farthest point sampling (T-FPS) to yield landmarks, for the efficient computation of diffusion geometry.
We consider the effectiveness and efficiency of this scheme relative to other possible landmark selection techniques:
\begin{itemize}
	\item \textbf{Random.} We select landmarks at random, as performed in~\cite{chen2011large}. Note that this preserves the density of the original manifold sampling.
	\item \textbf{Farthest Point Sampling (FPS).} We perform farthest point sampling with respect to the full particle path positions.
\end{itemize}

We evaluate the quality and efficiency of these techniques on the \textsf{Double Gyre} dataset. We integrate $7,200$ particles from a set of seeds
placed on a uniform grid of resolution $(120 \times 60)$, with a temporal resolution of $t=100$. We analyze the different landmark techniques as a function
of the number of selected landmarks. We perform evaluation by comparing the subspaces spanned by the
eigenvectors with largest eigenvalues to those of the ground truth subspace that uses all of the particles. Error is computed by projecting the landmark-based subspace onto the orthogonal
complement of the ground truth subspace, normalized by the ground truth subspace's Frobenius norm, i.e. if $U_l$ is the landmark subspace and $U$ is the ground truth subspace, the
error is: $\frac{(I-U U^{\intercal})U_l}{\lVert U \rVert_F}$.

\begin{figure}
	\begin{center}
		\includegraphics[width=\linewidth]{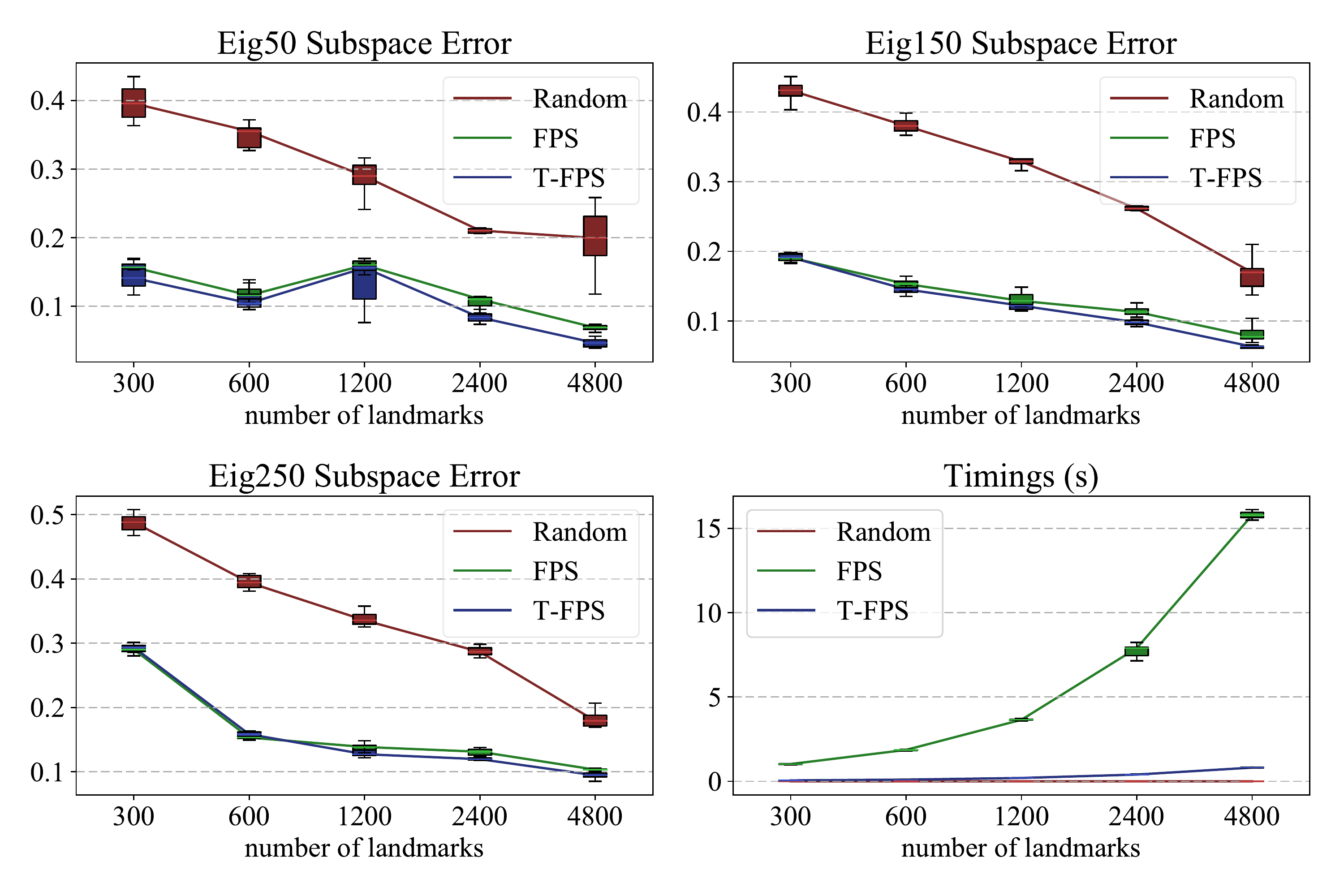}
	\end{center}
	\caption{We compare the effectiveness of different landmark selection schemes, for subspaces spanned by different amounts of eigenvectors. We find
	that T-FPS performs about as well as FPS and is significantly faster.}
	\label{fig:landmark_plots}
\end{figure}

We plot the errors and the computation times in Figure~\ref{fig:landmark_plots}, where each experiment is ran over 10 trials.
We plot error with respect to the eigenvector subspaces spanning the largest 50, 150, and 250 eigenvalues.
Multiple subspaces are considered to account for the fact that diffusion distances use different ranges of eigenvectors as a function of scale: the larger the scale, the more weight is placed
on the eigenvectors with largest eigenvalues. As shown, randomly sampling landmarks produces much higher error compared to the FPS-based techniques. Furthermore, there is
not much difference in error between using all time steps, and only using a temporal subsampling. However, the computational expense of using FPS is much higher than T-FPS.
Relative to our experiments shown in Table~\ref{table}, T-FPS ranges from approximately 5x to 20x faster than FPS, depending on the number of time steps $T$.
However, due to the $O(n \cdot n_l)$ complexity T-FPS does not necessarily scale well as the number of landmarks increases, as shown in the timings of Table~\ref{table}.

We also note that the error does not necessarily monotonically decrease, since the bandwidth is a function of the landmark distribution, and hence only captures
features at the scale of the landmarks. Nevertheless, in general the error does decrease as we increase the number of landmarks. % blah blah features about DG?
Similar experiments with other datasets produced consistent results with the Double Gyre, indicating the generalization of T-FPS.

\section{Discussion}

% conclusions
We have demonstrated how to use diffusion distances defined on particles for analyzing flow data. Diffusion distances allow us
to compute separation and similarity measures directly on particles, and in a multi-scale manner, where the scale can provide
the user an interface to understanding particle strength in separation/similarity. We have shown applicability to a wide number of applications,
as well as multiple sources of particle data, either integrated through time-varying vector fields or computed as particle-based simulation.

% limitations - complexity
Our method has several limitations that we intend to address in future work. First, we find that the computational complexity can become quite large as the
number of particles exceeds 1M, and the number of landmarks similarly grows large, i.e. 25K landmarks, as discussed in Section~\ref{sec:landeval}.
We intend to consider different types of sampling schemes that are more
efficient while still being effective. For instance, one option is to consider blue noise sampling, in particular the bilateral blue noise sampling technique~\cite{chen2013bilateral}.
Treating particle positions at the initial time step as the spatial domain, and the positions defined over the rest of the time steps as features,
such an approach is applicable to our setting, and should be much more efficient.

% limitations - scale
Currently, we leave scale as a user-defined parameter that can be adjusted for purposes of exploration. However, for certain applications exploration can be tedious when the proper scale
only exists in a certain set of small ranges. We will explore automatic techniques for scale selection, using the data to determine the sets of scales that are suitable for analysis.

% limitations - time
Another limitation of our approach is the restriction of particles to exist in the same temporal window, and to have positions defined at every time step.
We intend to explore techniques that allow for the comparison of particles for arbitrary time spans. In this manner, we can have a means of comparing
particles that just overlap in a certain time range, or potentially fail to overlap at all. Prior work in spectral clustering~\cite{brox2010object} has shown
the utility of this in analyzing trajectory data over arbitrary time intervals, but the interpretation with respect to diffusion remains unclear. 

% limitations - connectedness
Our analysis relies on particles to form a single connected component as defined by the matrix kernel $K$. However, certain types of particle-based simulations
such as fracture can produce data that leads to multiple connected components. Although we may process each component separately, diffusion distances between particles on
different components will be undefined. For future work we will investigate techniques to address this limitation.

% diffusion-based features
For future work we intend to explore different uses of diffusion for flow data. In particular, in the shape analysis community it is common to derive features 
from the heat kernel~\cite{sun2009concise}, a geometric object that is closely related to diffusion geometry. In particular, the heat kernel signature has proven a useful feature for multiple tasks. We think that leveraging these features could be useful for analyzing
flow, particularly for relating flow data produced from an ensemble of simulations.

\section*{Acknowledgements}

This work was partially supported by the National Science Foundation IIS-1654221. The \textsf{Cloud Collapse} and \textsf{Dust Settling} simulations were provided
by the Open Research Exeter database at the University of Exeter. We thank Matthew R. Bate and Daniel J. Price for their assistance in processing the simulation data.

\bibliographystyle{IEEEtran}
\bibliography{IEEEabrv,simflow}

\begin{thebibliography}{10}
\providecommand{\url}[1]{#1}
\csname url@samestyle\endcsname
\providecommand{\newblock}{\relax}
\providecommand{\bibinfo}[2]{#2}
\providecommand{\BIBentrySTDinterwordspacing}{\spaceskip=0pt\relax}
\providecommand{\BIBentryALTinterwordstretchfactor}{4}
\providecommand{\BIBentryALTinterwordspacing}{\spaceskip=\fontdimen2\font plus
\BIBentryALTinterwordstretchfactor\fontdimen3\font minus
  \fontdimen4\font\relax}
\providecommand{\BIBforeignlanguage}[2]{{%
\expandafter\ifx\csname l@#1\endcsname\relax
\typeout{** WARNING: IEEEtran.bst: No hyphenation pattern has been}%
\typeout{** loaded for the language `#1'. Using the pattern for}%
\typeout{** the default language instead.}%
\else
\language=\csname l@#1\endcsname
\fi
#2}}
\providecommand{\BIBdecl}{\relax}
\BIBdecl

\bibitem{lewis2015smoothed}
B.~T. Lewis, M.~R. Bate, and D.~J. Price, ``Smoothed particle
  magnetohydrodynamic simulations of protostellar outflows with misaligned
  magnetic field and rotation axes,'' \emph{Monthly Notices of the Royal
  Astronomical Society}, vol. 451, no.~1, pp. 288--299, 2015.

\bibitem{OttoKET12}
M.~Otto, A.~Kuhn, W.~Engelke, and H.~Theisel, ``2011 {IEEE} visualization
  contest winner: Visualizing unsteady vortical behavior of a centrifugal
  pump,'' \emph{{IEEE} Computer Graphics and Applications}, vol.~32, no.~5, pp.
  12--19, 2012.

\bibitem{janicke2008automatic}
H.~J{\"a}nicke, M.~B{\"o}ttinger, X.~Tricoche, and G.~Scheuermann, ``Automatic
  detection and visualization of distinctive structures in 3d unsteady
  multi-fields,'' in \emph{Computer Graphics Forum}, vol.~27, no.~3.\hskip 1em
  plus 0.5em minus 0.4em\relax Wiley Online Library, 2008, pp. 767--774.

\bibitem{oeltze2016cluster}
S.~Oeltze-Jafra, J.~R. Cebral, G.~Janiga, and B.~Preim, ``Cluster analysis of
  vortical flow in simulations of cerebral aneurysm hemodynamics,'' \emph{IEEE
  transactions on visualization and computer graphics}, vol.~22, no.~1, pp.
  757--766, 2016.

\bibitem{van2012visualization}
R.~van Pelt, S.~S. A.~M. Jacobs, B.~M. ter Haar~Romeny, and A.~Vilanova,
  ``Visualization of 4d blood-flow fields by spatiotemporal hierarchical
  clustering,'' in \emph{Comput. Graph. Forum}, vol.~31, no.~3, 2012, pp.
  1065--1074.

\bibitem{garth2007efficient}
C.~Garth, F.~Gerhardt, X.~Tricoche, and H.~Hans, ``Efficient computation and
  visualization of coherent structures in fluid flow applications,'' \emph{IEEE
  Transactions on Visualization and Computer Graphics}, vol.~13, no.~6, pp.
  1464--1471, 2007.

\bibitem{haller2000lagrangian}
G.~Haller and G.~Yuan, ``Lagrangian coherent structures and mixing in
  two-dimensional turbulence,'' \emph{Physica D: Nonlinear Phenomena}, vol.
  147, no.~3, pp. 352--370, 2000.

\bibitem{shadden2005definition}
S.~C. Shadden, F.~Lekien, and J.~E. Marsden, ``Definition and properties of
  lagrangian coherent structures from finite-time lyapunov exponents in
  two-dimensional aperiodic flows,'' \emph{Physica D: Nonlinear Phenomena},
  vol. 212, no.~3, pp. 271--304, 2005.

\bibitem{hong2014flda}
F.~Hong, C.~Lai, H.~Guo, E.~Shen, X.~Yuan, and S.~Li, ``Flda: latent dirichlet
  allocation based unsteady flow analysis,'' \emph{IEEE transactions on
  visualization and computer graphics}, vol.~20, no.~12, pp. 2545--2554, 2014.

\bibitem{pobitzer2012statistics}
A.~Pobitzer, A.~Le{\v{z}}, K.~Matkovi{\'c}, and H.~Hauser, ``A statistics-based
  dimension reduction of the space of path line attributes for interactive
  visual flow analysis,'' in \emph{Visualization Symposium (PacificVis), 2012
  IEEE Pacific}.\hskip 1em plus 0.5em minus 0.4em\relax IEEE, 2012, pp.
  113--120.

\bibitem{price2012smoothed}
D.~J. Price, ``Smoothed particle hydrodynamics and magnetohydrodynamics,''
  \emph{Journal of Computational Physics}, vol. 231, no.~3, pp. 759--794, 2012.

\bibitem{wu2012hierarchical}
J.~Wu, Z.~Lan, X.~Xiong, N.~Y. Gnedin, and A.~V. Kravtsov, ``Hierarchical task
  mapping of cell-based amr cosmology simulations,'' in \emph{Proceedings of
  the International Conference on High Performance Computing, Networking,
  Storage and Analysis}.\hskip 1em plus 0.5em minus 0.4em\relax IEEE Computer
  Society Press, 2012, p.~75.

\bibitem{ChandlerOJ15}
J.~Chandler, H.~Obermaier, and K.~I. Joy, ``Interpolation-based pathline
  tracing in particle-based flow visualization,'' \emph{{IEEE} Trans. Vis.
  Comput. Graph.}, vol.~21, no.~1, pp. 68--80, 2015.

\bibitem{CBJ16}
J.~Chandler, R.~Bujack, and K.~I. Joy, ``Analysis of error in
  interpolation-based pathline tracing,'' in \emph{EuroVis}, 2016.

\bibitem{ng2001spectral}
A.~Y. Ng, M.~I. Jordan, Y.~Weiss \emph{et~al.}, ``On spectral clustering:
  Analysis and an algorithm,'' in \emph{NIPS}, vol.~14, no.~2, 2001, pp.
  849--856.

\bibitem{HelmanH90}
J.~Helman and L.~Hesselink, ``Surface representations of two- and
  three-dimensional fluid flow topology,'' in \emph{{IEEE} Visualization},
  1990, pp. 6--13.

\bibitem{laramee2007topology}
R.~Laramee, H.~Hauser, L.~Zhao, and F.~Post, ``Topology-based flow
  visualization, the state of the art,'' \emph{Topology-based methods in
  visualization}, pp. 1--19, 2007.

\bibitem{garth2004tracking}
C.~Garth, X.~Tricoche, and G.~Scheuermann, ``Tracking of vector field
  singularities in unstructured 3d time-dependent datasets,'' in
  \emph{Visualization, 2004. IEEE}.\hskip 1em plus 0.5em minus 0.4em\relax
  IEEE, 2004, pp. 329--336.

\bibitem{FuchsKSWSHP10}
R.~Fuchs, J.~Kemmler, B.~Schindler, J.~Waser, F.~Sadlo, H.~Hauser, and
  R.~Peikert, ``Toward a lagrangian vector field topology,'' \emph{Comput.
  Graph. Forum}, vol.~29, no.~3, pp. 1163--1172, 2010.

\bibitem{haller2001distinguished}
G.~Haller, ``Distinguished material surfaces and coherent structures in
  three-dimensional fluid flows,'' \emph{Physica D: Nonlinear Phenomena}, vol.
  149, no.~4, pp. 248--277, 2001.

\bibitem{machado2016space}
G.~Machado, S.~Boblest, T.~Ertl, and F.~Sadlo, ``Space-time bifurcation lines
  for extraction of 2d lagrangian coherent structures,'' in \emph{Computer
  Graphics Forum}, vol.~35, no.~3.\hskip 1em plus 0.5em minus 0.4em\relax Wiley
  Online Library, 2016, pp. 91--100.

\bibitem{sadlo2011time}
F.~Sadlo, A.~Rigazzi, and R.~Peikert, ``Time-dependent visualization of
  lagrangian coherent structures by grid advection,'' in \emph{Topological
  Methods in Data Analysis and Visualization}.\hskip 1em plus 0.5em minus
  0.4em\relax Springer, 2011, pp. 151--165.

\bibitem{sadlo2007efficient}
F.~Sadlo and R.~Peikert, ``Efficient visualization of lagrangian coherent
  structures by filtered amr ridge extraction,'' \emph{IEEE Transactions on
  Visualization and Computer Graphics}, vol.~13, no.~6, pp. 1456--1463, 2007.

\bibitem{agranovsky2011extracting}
A.~Agranovsky, C.~Garth, and K.~I. Joy, ``Extracting flow structures using
  sparse particles.'' in \emph{VMV}, 2011, pp. 153--160.

\bibitem{RosslT12}
C.~R{\"{o}}ssl and H.~Theisel, ``Streamline embedding for {3D} vector field
  exploration,'' \emph{{IEEE} Trans. Vis. Comput. Graph.}, vol.~18, no.~3, pp.
  407--420, 2012.

\bibitem{LuCLSW13}
K.~Lu, A.~Chaudhuri, T.~Lee, H.~Shen, and P.~C. Wong, ``Exploring vector fields
  with distribution-based streamline analysis,'' in \emph{{IEEE} Pacific
  Visualization Symposium}, 2013, pp. 257--264.

\bibitem{ChaudhuriLSW14}
A.~Chaudhuri, T.~Lee, H.~Shen, and R.~Wenger, ``Exploring flow fields using
  space-filling analysis of streamlines,'' \emph{{IEEE} Trans. Vis. Comput.
  Graph.}, vol.~20, no.~10, pp. 1392--1404, 2014.

\bibitem{wang2014pattern}
Z.~Wang, J.~M. Esturo, H.-P. Seidel, and T.~Weinkauf, ``{Pattern Search in
  Flows based on Similarity of Stream Line Segments},'' in \emph{Vision,
  Modeling \& Visualization}.\hskip 1em plus 0.5em minus 0.4em\relax The
  Eurographics Association, 2014.

\bibitem{tao2016vocabulary}
J.~Tao, C.~Wang, C.-K. Shene, and R.~A. Shaw, ``A vocabulary approach to
  partial streamline matching and exploratory flow visualization,'' \emph{IEEE
  transactions on visualization and computer graphics}, vol.~22, no.~5, pp.
  1503--1516, 2016.

\bibitem{chen2007similarity}
Y.~Chen, J.~Cohen, and J.~Krolik, ``Similarity-guided streamline placement with
  error evaluation,'' \emph{IEEE Transactions on Visualization and Computer
  Graphics}, vol.~13, no.~6, pp. 1448--1455, 2007.

\bibitem{mcloughlin2013similarity}
T.~McLoughlin, M.~W. Jones, R.~S. Laramee, R.~Malki, I.~Masters, and C.~D.
  Hansen, ``Similarity measures for enhancing interactive streamline seeding,''
  \emph{IEEE Transactions on Visualization and Computer Graphics}, vol.~19,
  no.~8, pp. 1342--1353, 2013.

\bibitem{guo2014scalable}
H.~Guo, F.~Hong, Q.~Shu, J.~Zhang, J.~Huang, and X.~Yuan, ``Scalable
  lagrangian-based attribute space projection for multivariate unsteady flow
  data,'' in \emph{2014 IEEE Pacific Visualization Symposium}.\hskip 1em plus
  0.5em minus 0.4em\relax IEEE, 2014, pp. 33--40.

\bibitem{agranovsky2014improved}
A.~Agranovsky, D.~Camp, C.~Garth, E.~W. Bethel, K.~I. Joy, and H.~Childs,
  ``Improved post hoc flow analysis via lagrangian representations,'' in
  \emph{Large Data Analysis and Visualization (LDAV), 2014 IEEE 4th Symposium
  on}.\hskip 1em plus 0.5em minus 0.4em\relax IEEE, 2014, pp. 67--75.

\bibitem{bujack2015lagrangian}
R.~Bujack and K.~I. Joy, ``Lagrangian representations of flow fields with
  parameter curves,'' in \emph{Large Data Analysis and Visualization (LDAV),
  2015 IEEE 5th Symposium on}.\hskip 1em plus 0.5em minus 0.4em\relax IEEE,
  2015, pp. 41--48.

\bibitem{kuhn2014time}
A.~Kuhn, W.~Engelke, C.~R{\"o}ssl, M.~Hadwiger, and H.~Theisel, ``Time line
  cell tracking for the approximation of lagrangian coherent structures with
  subgrid accuracy,'' in \emph{Computer Graphics Forum}, vol.~33, no.~1.\hskip
  1em plus 0.5em minus 0.4em\relax Wiley Online Library, 2014, pp. 222--234.

\bibitem{shi2017analysis}
L.~Shi, L.~Zhang, W.~Cao, and G.~Chen, ``Analysis enhanced particle-based flow
  visualization,'' \emph{Electronic Imaging}, vol. 2017, no.~1, pp. 12--21,
  2017.

\bibitem{Froyland15}
G.~Froyland and K.~Padberg-Gehle, ``A rough-and-ready cluster-based approach
  for extracting finite-time coherent sets from sparse and incomplete
  trajectory data,'' \emph{Chaos}, vol.~25, no.~8, 2015.

\bibitem{Hadjighasem16}
A.~Hadjighasem, D.~Karrasch, H.~Teramoto, and G.~Haller, ``Spectral-clustering
  approach to lagrangian vortex detection,'' \emph{Phys. Rev. E}, vol.~93, p.
  063107, Jun 2016.

\bibitem{froyland2016dynamic}
G.~Froyland and E.~Kwok, ``A dynamic laplacian for identifying lagrangian
  coherent structures on weighted riemannian manifolds,'' \emph{arXiv preprint
  arXiv:1610.01128}, 2016.

\bibitem{karrasch2016geometric}
D.~Karrasch and J.~Keller, ``A geometric heat-flow theory of lagrangian
  coherent structures,'' \emph{arXiv preprint arXiv:1608.05598}, 2016.

\bibitem{banisch2017understanding}
R.~Banisch and P.~Koltai, ``Understanding the geometry of transport: Diffusion
  maps for lagrangian trajectory data unravel coherent sets,'' \emph{Chaos: An
  Interdisciplinary Journal of Nonlinear Science}, vol.~27, no.~3, p. 035804,
  2017.

\bibitem{lindeberg2013scale}
T.~Lindeberg, \emph{Scale-space theory in computer vision}.\hskip 1em plus
  0.5em minus 0.4em\relax Springer Science \& Business Media, 2013, vol. 256.

\bibitem{bauer2002vortex}
D.~Bauer and R.~Peikert, ``Vortex tracking in scale-space,'' in
  \emph{Proceedings of the symposium on Data Visualisation 2002}.\hskip 1em
  plus 0.5em minus 0.4em\relax Eurographics Association, 2002, pp. 233--ff.

\bibitem{fuchs2012scale}
R.~Fuchs, B.~Schindler, and R.~Peikert, ``Scale-space approaches to ftle
  ridges,'' in \emph{Topological Methods in Data Analysis and Visualization
  II}.\hskip 1em plus 0.5em minus 0.4em\relax Springer, 2012, pp. 283--296.

\bibitem{coifman2006diffusion}
R.~R. Coifman and S.~Lafon, ``Diffusion maps,'' \emph{Applied and computational
  harmonic analysis}, vol.~21, no.~1, pp. 5--30, 2006.

\bibitem{de2008hierarchical}
F.~De~Goes, S.~Goldenstein, and L.~Velho, ``A hierarchical segmentation of
  articulated bodies,'' in \emph{Computer graphics forum}, vol.~27,
  no.~5.\hskip 1em plus 0.5em minus 0.4em\relax Wiley Online Library, 2008, pp.
  1349--1356.

\bibitem{bronstein2010gromov}
A.~M. Bronstein, M.~M. Bronstein, R.~Kimmel, M.~Mahmoudi, and G.~Sapiro, ``A
  gromov-hausdorff framework with diffusion geometry for topologically-robust
  non-rigid shape matching,'' \emph{International Journal of Computer Vision},
  vol.~89, no. 2-3, pp. 266--286, 2010.

\bibitem{chen2011large}
X.~Chen and D.~Cai, ``Large scale spectral clustering with landmark-based
  representation.'' in \emph{AAAI}, 2011.

\bibitem{kuhn2012benchmark}
A.~Kuhn, C.~R{\"o}ssl, T.~Weinkauf, and H.~Theisel, ``A benchmark for
  evaluating ftle computations,'' in \emph{{IEEE} Pacific Visualization
  Symposium}, 2012, pp. 121--128.

\bibitem{sun2009concise}
J.~Sun, M.~Ovsjanikov, and L.~Guibas, ``A concise and provably informative
  multi-scale signature based on heat diffusion,'' in \emph{Computer graphics
  forum}, vol.~28, no.~5.\hskip 1em plus 0.5em minus 0.4em\relax Wiley Online
  Library, 2009, pp. 1383--1392.

\bibitem{singer2008non}
A.~Singer and R.~R. Coifman, ``Non-linear independent component analysis with
  diffusion maps,'' \emph{Applied and Computational Harmonic Analysis},
  vol.~25, no.~2, pp. 226--239, 2008.

\bibitem{kanaris2011three}
N.~Kanaris, D.~Grigoriadis, and S.~Kassinos, ``Three dimensional flow around a
  circular cylinder confined in a plane channel,'' \emph{Physics of Fluids},
  vol.~23, no.~6, p. 064106, 2011.

\bibitem{popinet2003gerris}
S.~Popinet, ``Gerris: a tree-based adaptive solver for the incompressible euler
  equations in complex geometries,'' \emph{Journal of Computational Physics},
  vol. 190, no.~2, pp. 572--600, 2003.

\bibitem{nadler2007fundamental}
B.~Nadler and M.~Galun, ``Fundamental limitations of spectral clustering,'' in
  \emph{Advances in neural information processing systems}, 2007, pp.
  1017--1024.

\bibitem{haller2005objective}
G.~Haller, ``An objective definition of a vortex,'' \emph{Journal of fluid
  mechanics}, vol. 525, pp. 1--26, 2005.

\bibitem{gunther2017generic}
T.~G{\"u}nther, M.~Gross, and H.~Theisel, ``Generic objective vortices for flow
  visualization,'' \emph{ACM Transactions on Graphics (TOG)}, vol.~36, no.~4,
  p. 141, 2017.

\bibitem{loren2015toroidal}
P.~Lor{\'e}n-Aguilar and M.~R. Bate, ``Toroidal vortices and the conglomeration
  of dust into rings in protoplanetary discs,'' \emph{Monthly Notices of the
  Royal Astronomical Society: Letters}, vol. 453, no.~1, pp. L78--L82, 2015.

\bibitem{skillman2014dark}
S.~W. Skillman, M.~S. Warren, M.~J. Turk, R.~H. Wechsler, D.~E. Holz, and
  P.~Sutter, ``Dark sky simulations: Early data release,'' \emph{arXiv preprint
  arXiv:1407.2600}, 2014.

\bibitem{schatz2016interactive}
K.~Schatz, C.~M{\"u}ller, M.~Krone, J.~Schneider, G.~Reina, and T.~Ertl,
  ``Interactive visual exploration of a trillion particles,'' in \emph{Large
  Data Analysis and Visualization (LDAV), 2016 IEEE 6th Symposium on}.\hskip
  1em plus 0.5em minus 0.4em\relax IEEE, 2016, pp. 56--64.

\bibitem{shivashankar2016felix}
N.~Shivashankar, P.~Pranav, V.~Natarajan, R.~van~de Weygaert, E.~P. Bos, and
  S.~Rieder, ``Felix: A topology based framework for visual exploration of
  cosmic filaments,'' \emph{IEEE Transactions on Visualization and Computer
  Graphics}, vol.~22, no.~6, pp. 1745--1759, 2016.

\bibitem{warren_2014_10777}
\BIBentryALTinterwordspacing
M.~S. Warren, A.~Friedland, D.~E. Holz, S.~W. Skillman, P.~M. Sutter, M.~J.
  Turk, and R.~H. Wechsler, ``Dark sky simulations collaboration,'' Jul. 2014.
  [Online]. Available:
  \url{http://darksky.slac.stanford.edu/scivis2015/data.html}
\BIBentrySTDinterwordspacing

\bibitem{biau2011weighted}
G.~Biau, F.~Chazal, D.~Cohen-Steiner, L.~Devroye, C.~Rodriguez \emph{et~al.},
  ``A weighted k-nearest neighbor density estimate for geometric inference,''
  \emph{Electronic Journal of Statistics}, vol.~5, pp. 204--237, 2011.

\bibitem{chen2013bilateral}
J.~Chen, X.~Ge, L.-Y. Wei, B.~Wang, Y.~Wang, H.~Wang, Y.~Fei, K.-L. Qian, J.-H.
  Yong, and W.~Wang, ``Bilateral blue noise sampling,'' \emph{ACM Transactions
  on Graphics (TOG)}, vol.~32, no.~6, p. 216, 2013.

\bibitem{brox2010object}
T.~Brox and J.~Malik, ``Object segmentation by long term analysis of point
  trajectories,'' \emph{Computer Vision--ECCV 2010}, pp. 282--295, 2010.

\end{thebibliography}

\end{document}